\documentclass[traditabstract]{aa} 

\usepackage{graphicx}
\usepackage{txfonts}
\usepackage{subfigure}
\usepackage{txfonts}
\usepackage{lscape}
\usepackage{rotating}

\begin{document}

\newcommand{\be}{\begin{equation}}
\newcommand{\ee}{\end{equation}}
\newcommand{\cmq}{cm$^{-3}$}
\newcommand{\um}{$\mu m$}
\newcommand{\sori}{$\sigma$~Ori}
\newcommand{\roph}{$\rho$-Oph}
\newcommand{\Msun}{M$_\odot$}
\newcommand{\Lsun}{L$_\odot$}
\newcommand{\Ha}{H$\alpha$}
\newcommand{\Pab}{Pa$\beta$}
\newcommand{\Pag}{Pa$\gamma$}
\newcommand{\Brg}{Br$\gamma$}
\newcommand{\Lpag}{L(Pa$\gamma$)}
\newcommand{\Teff}{T$_{eff}$}
\newcommand{\Lstar}{$L_{*}$}
\newcommand{\Rstar}{$R_{*}$}
\newcommand{\Mstar}{$M_{*}$}
\newcommand{\Laccc}{$L_{acc,c}$}
\newcommand{\Laccl}{$L_{acc,l}$}
\newcommand{\Lacc}{$L_{acc}$}
\newcommand{\Maccc}{$\dot M_{acc,c}$}
\newcommand{\Maccl}{$\dot M_{acc,l}$}
\newcommand{\Macc}{$\dot M_{acc}$}
\newcommand{\Mloss}{$\dot M_{loss}$}
\newcommand{\Mwind}{$\dot M_{wind}$}
\newcommand{\Ldisk}{L$_{disk}$}
\newcommand{\Myr}{M$_\odot$/yr}

   \title{X-shooter spectroscopy of young stellar objects: \\
I --  Mass accretion rates of low-mass T Tauri stars in $\sigma$~Orionis 
   \thanks{
Based on observations collected at the European Southern Observatory, Chile.
Program 084.C-0269(A), 086.C-0173(A)}}

\titlerunning{Mass accretion rates of low-mass T Tauri stars}

   \author{E. Rigliaco\inst{1,2},
          A. Natta\inst{1,5}, L.Testi\inst{1,6}, S. Randich\inst{1},  J.M. Alcal\`a\inst{3}, E. Covino\inst{3},  B. Stelzer\inst{4}  
          }

   \institute {INAF/Osservatorio Astrofisico of Arcetri, Largo E. Fermi, 5, 50125 Firenze, Italy\\
              \email{rigliaco@lpl.arizona.edu} \\
         \and
   {Department of Planetary Science, Lunar and Planetary Lab, University of Arizona, 1629, E. University Blvd, 85719, Tucson, AZ, USA}\\	
\and 
 {INAF/Osservatorio Astronomico di Capodimonte, Salita Moiariello, 16  80131, Napoli, Italy}\\
	\and 
   {INAF/Osservatorio Astronomico di Palermo, Piazza del Parlamento 1, 90134 Palermo, Italy }\\
	\and
   {DIAS/Dublin Institute for Advanced Studies, Burlington Road, Dublin 4, Ireland}\\
\and 
   {ESO/European Southern Observatory, Karl-Schwarzschild-Strasse 2
D-85748 Garching bei M\"unchen, Germany}
             }

	\offprints{rigliaco@lpl.arizona.edu}
   \date{Received ...; accepted ...}

 
  \abstract
   {
We present high-quality, medium resolution X-shooter/VLT spectra in the range 300-2500 nm for a sample 
of 12  very low-mass stars in the $\sigma$ Orionis cluster.  
The sample includes eight stars with evidence of disks from {\it Spitzer} and four without, 
with masses ranging from 0.08 to 0.3 M$_\odot$. 
The aim of this first paper is to investigate the reliability of the many accretion tracers  currently used 
to measure the mass accretion rate in low-mass, young stars and the accuracy of the correlations 
between these secondary tracers (mainly accretion line luminosities) found in the literature. 
We use our  spectra to measure the accretion luminosity from the continuum excess emission in the UV and visual;
the derived mass accretion rates range from 10$^{-9}$ M$_{\odot}$ yr$^{-1}$ 
down to 5$\times$10$^{-11}$ M$_{\odot}$ yr$^{-1}$, allowing us to investigate the 
behavior of the accretion-driven emission lines in very-low mass accretion rate regimes.
We compute the luminosity of ten accretion-driven emission lines, from the UV to the near-IR, all obtained simultaneously. 
In general, most of the  secondary tracers correlate well with the accretion luminosity derived from the continuum excess emission. 
We recompute the relationships between the accretion luminosities and the line luminosities,  
we confirm the validity of the correlations given in the literature, with the possible exception of \Ha. 
Metallic lines, such as the CaII IR triplet or the Na I line at 589.3 nm, show a larger dispersion. 
When looking at individual objects, we find that the Hydrogen recombination lines, 
from the UV to the near-IR, give good and consistent measurements of \Lacc\, 
often in better agreement than the uncertainties introduced by the adopted correlations. 
The average \Lacc\ derived from several Hydrogen lines, measured simultaneously, have a much reduced error.
This suggests that some of the spread in the literature correlations may be  due to the  use of non-simultaneous observations of lines and continuum. 
Three stars  in our sample deviate from this behavior, and we discuss them individually.}



   \maketitle
%

\section{Introduction}\label{introduction}

Accretion of matter onto T Tauri stars is an important aspect of the star formation process, 
and plays a fundamental role in shaping the structure and evolution of proto-planetary disks. 
Magnetospheric accretion (Uchida \& Shibata~1985, K\"oenigl~1991, Shu et al.~1994) 
is the accepted paradigm that explains the accretion in Classical T Tauri stars (CTTSs) and lower-mass objects. 
In this family of models, material from the inner edge of the disk is channelled through the stellar magnetosphere field lines and flows onto the star. 

In a CTTS it is possible to identify two different regions in the magnetospheric accretion model structure: 
the accretion columns and the hot spots on the stellar surface. 
In the accretion columns there is gas which is accreting onto the star channelled along the 
magnetic field lines, 
while the hot spots are produced where the infalling material impacts onto the stellar surface.
The accretion luminosity is released as continuum and line emission, formed in the hot spots 
and/or in the accreting gas columns.

The intensity of the continuum emission, its wavelength dependence, the intensity and profiles of the various emission lines can be used to derive quantitative information on the accretion process itself (e.g., Hartmann et al. 2006). In particular, the integrated continuum and line luminosity \Lacc\ is proportional to the mass accretion rate, which can be computed from it once the ratio of the stellar mass to radius is known. Measurements of \Lacc\ require well-calibrated observations of the continuum flux over a large range of wavelengths, as well as an adequate knowledge of the photospheric and chromospheric spectrum of the star, which need to be subtracted from the observed flux to isolate the accretion emission. This has been possible for a relatively small sample of objects only (Hartigan et al. 1995, Muzerolle et al.~2003, Hartigan et al. 1991; Valenti et al. 1993; Calvet \& Gullbring 1998; Gullbring et al. 1998, Herczeg \& Hillenbrand.~2008, hereafter HH08).  
However, it has been noticed that in these objects the luminosity or flux of several lines correlates quite well with \Lacc. Using these correlations, it has been possible to estimate \Lacc\, and therefore \Macc\, in a large number of young stars.
These {\it secondary indicators} span a huge range of wavelength. 
In the far-ultraviolet (FUV) Herczeg et al. (2002) and Yang et al. (2012) find that the 
OI$\lambda$103.4 nm triplet, SiIV$\lambda$139.4,140.3 nm doublet and CIV$\lambda$154.9 nm doublet 
are tightly correlated with \Lacc. In the soft X-rays, Telleschi et al. (2007) and Curran et al. (2011) 
find that the low-density plasma component in the post-shock region correlates with \Macc. 
From the near-UV to the near-IR several hydrogen 
recombination lines can be used to estimate  
\Lacc\ (\Ha, H$\beta$, H11, Pa$\beta$, Pa$\gamma$), as well as other lines 
CaII$\lambda$854.2 nm, CaII$\lambda$866.2 nm, HeI$\lambda$587.6 nm, NaI$\lambda$589.3 nm, as shown by 
Fang et al.~(2009); Mohanty et al.~(2005), HH08, Natta et al.~(2004), Gatti et al.~(2008) among many others.  
Also the excess emission in the U-band is well correlated with \Lacc\ (Gullbring et al. 1998), and can be used to measure it when the reddening is not large.
Finally a widely used accretion indicator is the \Ha\ 10\% line width, which has been found to be correlated with \Macc\ (Natta et al. 2004), although with large dispersion. 
The existence  of secondary indicators in different regions of the spectrum
is very important, as the more direct method of measuring \Lacc\ integrating the excess emission over the whole wavelength range is often impractical or altogether impossible. 
It is, therefore, necessary to estimate the relative reliability of the different indicators over the largest possible range of the parameters, i.e., \Lacc, \Macc, or \Mstar. \\
When several indicators are measured for the same object, the accretion rate may differ by  a factor of ten or more. Since in many cases different indicators are not observed simultaneously, one possible reason is the well-known time variability of  the pre-main sequence stars 
 (Joy, A. 1945, Herbst et al. 1994, Herbst et al. 2007). 
Up to now it has been difficult to observe several accretion tracers simultaneously, 
especially because they span from the UV wavelength region (e.g., the Balmer series lines) to the near-IR 
(e.g., Pa$\beta$, Pa$\gamma$ and Br$\gamma$ hydrogen recombination lines). 
This means that when comparing non simultaneous observations of these indicators 
an error due to the time variability can not be ruled out. 
This observational gap can be filled in with the use of broad spectral range spectrographs
like X-shooter@VLT. 
X-shooter is an ideal instrument for comparing different accretion tracers 
because it covers simultaneously  the wavelength range from $\sim$300 to $\sim$2500 nm.\\

We present here a systematic simultaneous study of primary 
and secondary accretion indicators in young stellar objects. Our target sample is composed of
young low-mass 
T Tauri stars in the $\sigma$ Orionis star forming region. $\sigma$ Orionis is located at a distance 
of $\sim$360 pc ({\it Hipparcos} distance 352$^{+166}_{-85}$pc, Brown et al. 1994, Perryman et al. 1997) 
and has an age of $\sim$3 Myr (Zapatero-Osorio et al. 2002, Fedele et al. 2010). The region has 
negligible extinction (B\'ejar et al. 1999, Oliveira et al. 2004), and has been extensively studied 
in the optical, X-rays, and infrared (e.g. Rigliaco et al. 2011a, Kenyon et al. 2005, Zapatero-Osorio et al. 2002, 
Jeffries et al. 2006, Franciosini et al. 2006, Hernandez et al. 2007). \\
The observations presented in this paper have been carried out 
 as part of a larger study of the accretion properties in very low mass stars and brown dwarfs 
in nearby star forming regions with X-shooter (Alcal\`a et al., 2011, Alcal\`a et al. in prep.). 

In this paper, we present results related to the accretion phenomena. 
Additional results about 
the properties of the non-accreting objects and wind phenomena will be shown in upcoming 
papers of Manara et al.~(2012, in prep.) and Rigliaco et al.~(in prep.), respectively. \\
The structure of the paper is as follows. Section 2 summarizes the details of the observations, data reduction and the instrument set-up. 
Section 3 describes the properties of the sample. In Section 4 we discuss how  the accretion luminosities and mass accretion rates are measured  
from the excess continuum emission. 
The measurement of \Lacc\ from the emission lines is described in Section 5. In Section 5.1 and Section 5.2 
we compare all the indicators observed 
simultaneously with X-Shooter, and investigate their reliability as accretion tracers.   
Notes on individual targets are given in Section 5.3. 
A summary can be found in Section 6. \\


\section{Observations and data reduction}

\small
   \begin{table}
   \centering
      \caption[]{Journal of the observations. 
The last column shows the identification name, as reported in Hernandez et al.~(2007).  }
         \label{table1_log}
    \begin{tabular}{c c c c c}
    	\hline
	RA & DEC & obs date & t{\it{exp}}   & Name \\
	   &     & (y-m-d)  &(sec)      &   		 \\
		\hline
05:38:13.18 & -02:26:8.629 & 2011-01-11 & 900$\times$2  & SO397 \\
05:38:23.58 & -02:20:47.47 & 2011-01-11 & 900$\times$4  & SO490 \\
05:38:25.41 & -02:42:41.15 & 2009-12-24 & 900$\times$6  & SO500 \\
05:38:34.04 & -02:36:37.33 & 2009-12-22 & 600$\times$2  & SO587 \\
05:38:38.57 & -02:41:55.79 & 2009-12-24 & 900$\times$4  & SO641 \\
05:38:39.01 & -02:45:31.97 & 2009-12-24 & 600$\times$2 & SO646 \\
05:38:54.91 & -02:28:58.19 & 2009-12-24 & 1200$\times$2 & SO797 \\
05:39:1.938 & -02:35:2.831 & 2009-12-22 & 900$\times$4 & SO848 \\
05:39:11.41 & -02:33:32.8 & 2009-12-22 & 900$\times$4 & SO925 \\
05:39:20.25 & -02:38:25.8 & 2009-12-25 & 1200$\times$2 & SO999 \\
05:39:53.63 & -02:33:42.88 & 2011-01-13 & 900$\times$2 & SO1260 \\
05:39:54.22 & -02:27:32.87 & 2011-01-12 & 900$\times$4 & SO1266 \\
16:07:49.59 & -39:04:28.79 & 2011-04-06 & 300$\times$2 & Sz94 \\
		\hline

	\end{tabular}
   \end{table}
\normalsize

The sample analyzed in this paper has been observed with X-shooter on VLT as part of the 
Italian INAF/GTO program on star forming regions (Alcal\`a et al.~2011). 
Observations were performed in visitor mode on 21, 23 and 24 December 2009 and 11-12 January 2011. 
We observed 12 objects, obtaining medium-resolution spectra 
covering $\sim$ 300-2500 nm range. 
The observations log (Table \ref{table1_log}) lists the coordinates of the targets, 
the exposure time, and the abbreviated object names used throughout this paper 
(the last row of this Table reports the information of  a class III star with spectral type M4, 
observed within the GTO program in the Lupus star forming region. 
This star has been used as class III template in this paper). 

The targets were observed in nodding mode with the $11''\times1.0''$ slit in the UVB-arm, 
and with $11''\times0.9''$ slit in the VIS and NIR-arms. 
This instrument configuration yields a resolution R $\sim$ 5100 
over the UVB-arm, which covers a wavelength range 300-590 nm, 
and the NIR-arm, which covers a wavelength range 1000-2480 nm,
while with the VIS-arm (wavelength range from $\sim$ 580 nm to $\sim$ 1000 nm) we achieved 
a resolution R $\sim$ 8800. \\
The data reduction was performed independently for each spectrograph arm 
using the X-shooter pipeline version 1.1.0, following the standard steps 
which include bias subtraction, flat-fielding, optimal extraction, wavelength 
calibration, sky subtraction. 
The extraction of the 1D spectra and the subsequent data analysis was performed 
in STARE mode for UVB and VIS-arms, 
and in NODDING mode for the NIR-arm. 
Flux calibration has been achieved using the context LONG within MIDAS\footnote{Munich Image Data Analysis System}. 
For this purpose, a response function was derived by interpolating 
the counts/standard-flux ratio of the flux standards (observed the same night as the objects, generally 
under photometric sky conditions), with a third-order spline function, after airmass correction. 
The flux-match of the three arms is excellent (Alcal\`a et al. 2011), 
and the flux-calibrated spectra show on average 
a very good agreement with the flux derived from photometric measurements available in the literature. 
Only in a few cases observed in not completely photometric conditions, we have adjusted the flux-calibrated spectra shifting the spectra on the photometric flux. 
The data were analyzed using the task SPLOT within 
the IRAF\footnote{{\sc iraf} is distributed by National Optical Astronomy
Observatories, which is operated by the Association of Universities
for Research in Astronomy, Inc., under cooperative agreement with the
National Science Foundation.} package.


\section{The sample}

The 12 targets for this study
have been selected among the sample observed in the U-Band by Rigliaco et al.~(2011a) 
with FORS1@VLT, covering a range of masses between $\sim$0.08-$\sim$0.3 \Msun. 
Objects with mid-infrared colors that suggest they are young stars 
possessing disks (class II), and sources without obvious IR excess emission (class III) to be used as template  
were selected.\\

Spectral types (SpT) were estimated using the various indices by Riddick et al. (2007). These authors provide calibrations specifically for young M dwarfs. The flux ratios were derived from the flux-calibrated optical spectra. The final SpT assigned to a given object was estimated as the average 
  SpT resulting from the various indices. 
The effective temperature $\rm{T_{eff}}$ was obtained for each star 
using the temperature scale of Luhman et al.~(2003).
Luminosities were then computed using the I-band magnitudes (as listed in Table~\ref{table_target_properties}), and the bolometric correction for 
zero age main sequence (ZAMS) stars as reported by Luhman et al.~(2003). 
As described in Sect.~\ref{introduction}, 
the distance of $\sigma$\,Ori is $360$\,pc and the extinction is negligible. 
The uncertainty in the SpT of about half a subclass  corresponds to a typical error of +/- 70K on the photospheric effective temperature.
The error on the luminosity is due to the adopted bolometric correction (which is linked to the error on $\rm{T_{eff}}$),  
to the uncertainty on the measurement of the I-band magnitude (assumed to be $\sim$0.2 mag), and to the uncertainty on the 
distance of the cluster which may vary from $\sim$330 pc (Caballero, J. 2008) to $\sim$ 470 pc (De Zeeuw et al. 1999). 
A detailed analysis of the error due to the uncertainty on the distance is done by Rigliaco et al.~(2011c). 
We estimate a typical uncertainty in luminosity of $\sim$0.2 dex.  
Based  on the location of the targets on the HR diagram and on a comparison with 
the Baraffe et al.~(1998) evolutionary tracks, we have estimated the object masses and ages. 
The properties of the stars in our sample are summarized in Table~\ref{table_target_properties}. 
The determinations of the radius and of the mass are also affected by the distance uncertainty. We assume a typical 
error on these parameters of $\sim$0.05 dex. \\

   \begin{table}
   \centering
      \caption[]{Stellar parameters and observed properties. }
         \label{table_target_properties}
    \begin{tabular}{c c c c c c c c}
    	\hline
	Name & SpT & $\rm{T_{eff}}$ & Lum & Radius & Mass & class & I \\
     &          & (K)  & (L$_{\odot}$) &	(R$_{\odot}$) & (M$_{\odot}$) & & mag  \\
		\hline
SO397 &	M4.5 & 3200 & 0.19 & 1.45 & 0.20 & II & 14.10\\
SO490 &	M5.5 & 3060 & 0.08 & 1.02 & 0.14 & II &15.32\\
SO500 &	M6 & 2990 & 0.02 & 0.47 & 0.08 & II &	17.30\\
SO587 &	M4.5 & 3200 & 0.28 & 1.73 & 0.20 	& II & 13.72\\
SO641 &	M5 & 3125 & 0.03 & 0.57 & 0.12 	& III  &16.36\\
SO646 &	M3.5 & 3350 & 0.10 & 0.97 & 0.30 & II	 &14.58\\
SO797 &	M4.5 & 3200 & 0.05 & 0.76 & 0.18  & III &	15.50\\
SO848 &	M4 & 3270 & 0.02 & 0.46 & 0.19 & II & 16.38\\
SO925 &	M5.5 & 3060 & 0.03 & 0.58 & 0.10 & III  &	16.54\\
SO999 &	M5.5 & 3060 & 0.06& 0.91 & 0.14 & III & 15.56\\
SO1260 & M4	& 3270 & 0.13 & 1.12 & 0.26 & II & 14.44\\
SO1266 & M4.5 & 3200 & 0.06 & 0.84 & 0.20 & II  &	15.30\\
		\hline
	\end{tabular}
   \end{table}

\begin{figure}
	\begin{center}
	
		\subfigure[SEDs of the class II objects]{
		\label{SED2}
		\includegraphics[width=9cm]{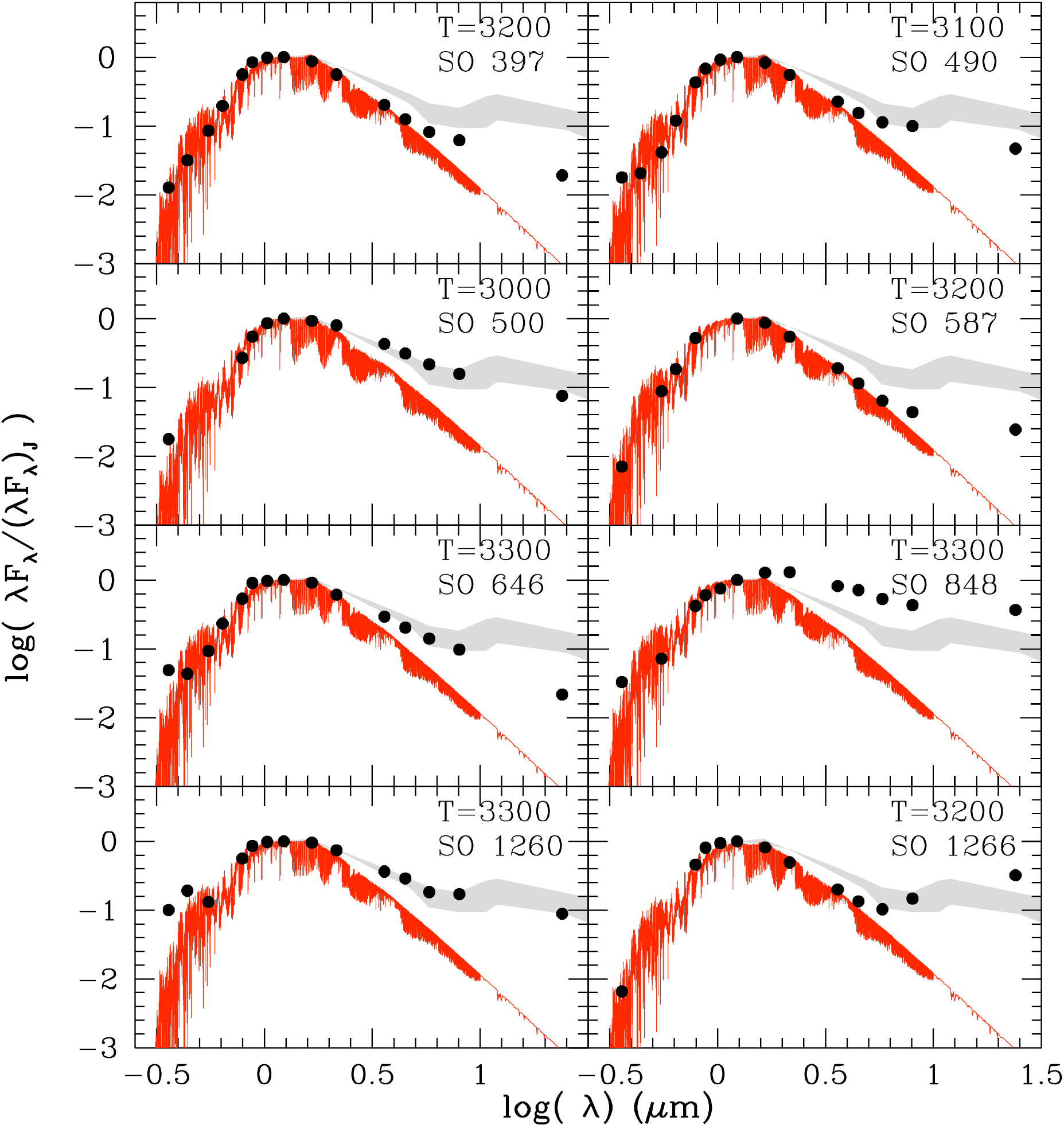}
		}

		\subfigure[SEDs of the class III objects]{
		\label{SED3}
		\includegraphics[width=9cm]{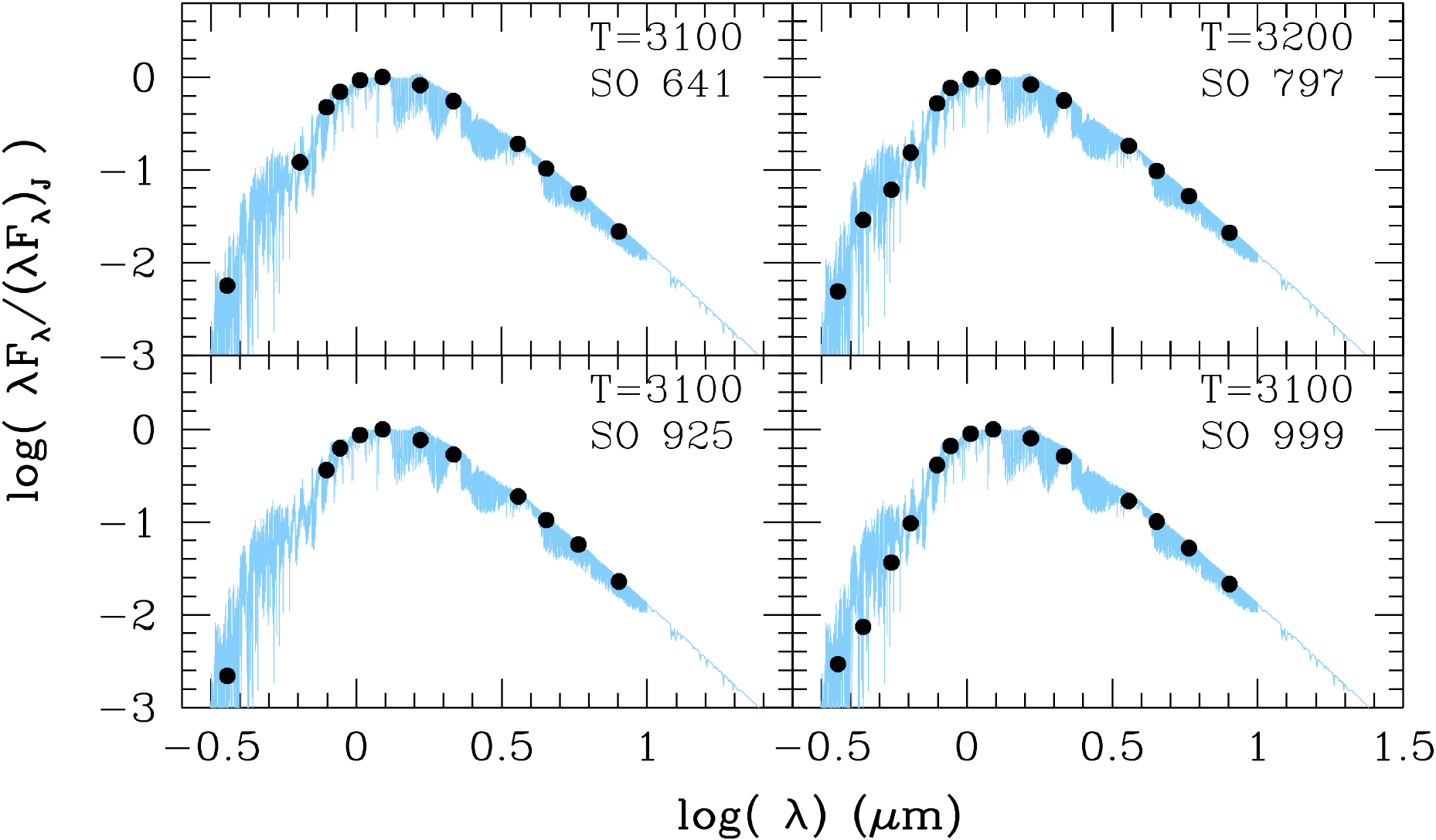}
		}
	\end{center}
	\caption{SEDs for the sample analyzed in this paper. 
Effective temperature of the adopted photospheric template is indicated as well as the name of the target. 
For all the stars with excess above the photosphere, the median disk SED in Taurus (grey region) is displayed, scaled to the J-band Flux. 
Black points are the photometric data (we do not apply any 
correction for the extinction being the reddening in the $\sigma$Ori star forming region negligible (B\'ejar et al. 1999, Oliveira et al. 2004)) . 
}
\label{SED_tot}
\end{figure}

As expected from previous photometric classification (Hernandez et al. 2007),
eight stars in our sample 
show a Spectral Energy Distribution (SED, in Fig.~\ref{SED_tot}) with IR excess emission, in particular
in the IRAC spectral range (from 3.6~$\mu$m to 8.0~$\mu$m), 
indicating the presence of an optically thick circumstellar disk, i.e. they are 
class\,II objects (see Fig.~\ref{SED_tot}(a)).
The remaining four stars in our sample show the 
typical color of the stellar photospheres and no IR excess. The SEDs of these 
class\,III objects are shown in Fig.~\ref{SED_tot}(b). 
 The SEDs shown in Fig.~\ref{SED_tot} are scaled to the J-band flux, while in the following 
figures the spectra are normalized to 700 nm.  Previous work on the veiling 
(Hartigan et al. 1995, White \& Ghez 2001, Fischer et al. 2011)
 show that there may be some excess continuum at and beyond this wavelength, 
however in our sample 
the veiling is small (see the following sections), allowing us to use the region around 700 nm to 
normalize the spectra.
In Fig.~\ref{SED_tot}, the black points are the available photometry, 
{\it U-band} from Rigliaco et al.~(2011a), {\it BVRI} from  the literature (see Rigliaco et al.~2011a),
 {\it JHK} from 2MASS photometry (Cutri et al.~2003), 
{\it z,Y} from  the UKIDSS survey, and {\it IRAC/MIPS} magnitudes from the 
{\it Spitzer} survey. 
In Fig.~\ref{SED_tot}(a), the red curves represent the photospheric contribution 
 obtained using the NextGen models 
(Hauschildt et al. 1999a, Hauschildt et al. 1999b, Allard et al. 2000),
normalized to the J-band, and for comparison purpose, 
we show, as grey region, the median SED for CTTSs in Taurus  
derived from previous ground-based and {\it IRAS} data (D'Alessio et al.~1999).

\section{Balmer and Paschen continua as accretion diagnostics}
\label{balmer_section}

The accretion luminosity is released as continuum and line  emission  over a very large range of wavelengths. In general, the continuum luminosity dominates over the emission in lines, which is generally neglected in the literature, although, as we will see in Sect.~\ref{line_contribution}, this may not be the case at low mass-accretion rate. 
We will define in the following \Laccc\ to be the continuum luminosity only. 

\begin{figure*}[ht]
   \centering
   \includegraphics[width=15cm]{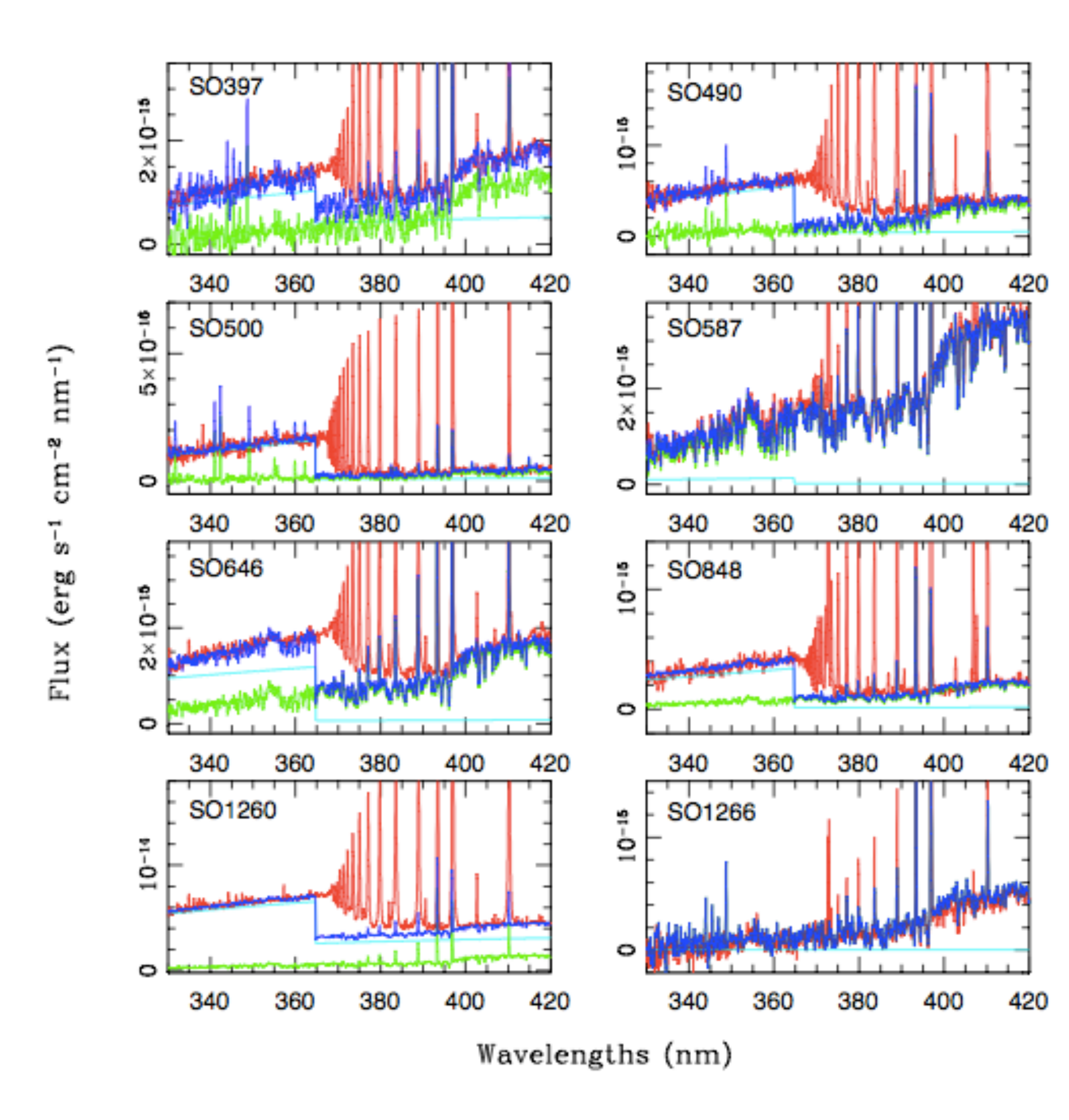}
   \caption{Comparison between the observed spectra and the fit obtained considering the slab model. 
Red spectra represent the observed emission in the region of the Balmer jump after a 
smoothing by boxcar seven. The green spectra show the photospheric template, the cyan 
 lines are the  synthetic accretion spectra from  the slab model. 
The sum of the slab model plus the template is shown by the blue spectra.}
              \label{jump_all}%
    \end{figure*}

The excess emission is clearly seen in the Balmer continuum, where the photospheric and chromospheric emission of the stars is very small. 
Of the eight class II stars in our sample, six show clear evidence of it, 
as seen in Fig.~\ref{jump_all}, where the spectrum of each star is compared to that of a class III object of similar spectral type, normalized at 700 nm (the adopted class III template stars for each class II object 
are listed in Table~\ref{frac_lum}). \\
Most stars show clearly the Balmer Jump at the Balmer edge
(note that the Balmer limit occurs at $\lambda$346.7 nm, but the line blending 
in the Balmer series shifts the apparent jump to 370 nm). 

In Table~\ref{frac_lum} we list the values of the observed Balmer Jump (BJ$_{obs}$), 
defined as  the ratio of the flux at 360 nm to the flux at 420 nm,
and of the intrinsic Balmer Jump (BJ$_{intr}$), measured after subtracting off the
photospheric template from the class II star. 
This value is generally computed after dereddening the spectra, however in the $\sigma$ Orionis 
star forming region the extinction is negligible (Oliveira et al. 2004). 
The observed balmer jump ranges between $\sim$0.2 
and $\sim$3.5 for the class II objects, as shown in Table~\ref{frac_lum}, column 3. 
BJ$_{obs}$ for the class III objects spans between $\sim$0.2 and $\sim$0.5. 
In two class II objects, namely SO587 and SO1266, BJ$_{obs}$ is in the same range of the class III stars. These objects will be discussed in the following of the paper.
The intrinsic balmer jump ranges between $\sim$1.4 and $\sim$14.4, 
as reported in Table~\ref{frac_lum}, column  4. 

The accretion continuum emission in the Paschen continuum 
(i.e., for 346 nm$ \lesssim \lambda \lesssim$820 nm) 
is usually a smaller fraction of the photospheric continuum than in the UV and it is more easily detected as 
veiling\footnote{The veiling $r_{\lambda}$ at wavelength $\lambda$ is defined as $(F_{obs}-F_{phot})$/$F_{phot}$ where $F_{obs}$ is the observed flux, and $F_{phot}$ is the photospheric flux.}  of the photospheric absorption lines  
which are filled-in by the excess continuum. \\
Figure 3 shows the CaI $\lambda$422.6 nm line for the eight class II stars in our sample. 
We find very little evidence (if any) of veiling in 6/8 objects. 
Only in two cases ( SO397 and SO1260) the CaI line is clearly filled-in. 
In some of our stars (SO490 and SO500), the CaI 422.6 nm line is slightly 
shallower than expected from a comparison to the non-accreting template. 
However, this is likely due to the overlapping emission of many lines. Moreover, 
the depth of this line may also be affected by stellar rotation 
(Mauas \& Falchi~1996, Short et al.~1997).\\

To compute \Laccc, we model the excess luminosity as the emission of a slab of hydrogen of fixed electron density ($N_e$), 
temperature ($T_e$) and length ($L$), under the assumption of local thermal equilibrium (LTE). 
Similar models, although very simple and not realistic,  have been used by Valenti et al. (1993) and HH08, among others, 
and have been found to reproduce well the wavelength dependence of the observed excess emission. 
These models allow us to account for the ultraviolet emission outside of the
observed range and to interpolate between the wavelength of the veiled lines. 
We add the emission of the slab to that of the adopted class III template, varying the model parameters
 until we find a satisfactory fit to the observed continuum
over the whole spectral range from UV to red. Note that there are five free parameters, namely the slab parameters ($T_e, N_e, L$), 
a scale factor for  the slab emission and the normalization constant of the class III template, 
which all need to be constrained simultaneously. 
We have varied the slab parameters over a large range of values, 
and adopted different class III templates whenever possible. 

\begin{figure}
   \centering
  \includegraphics[width=9.5cm, angle=0]{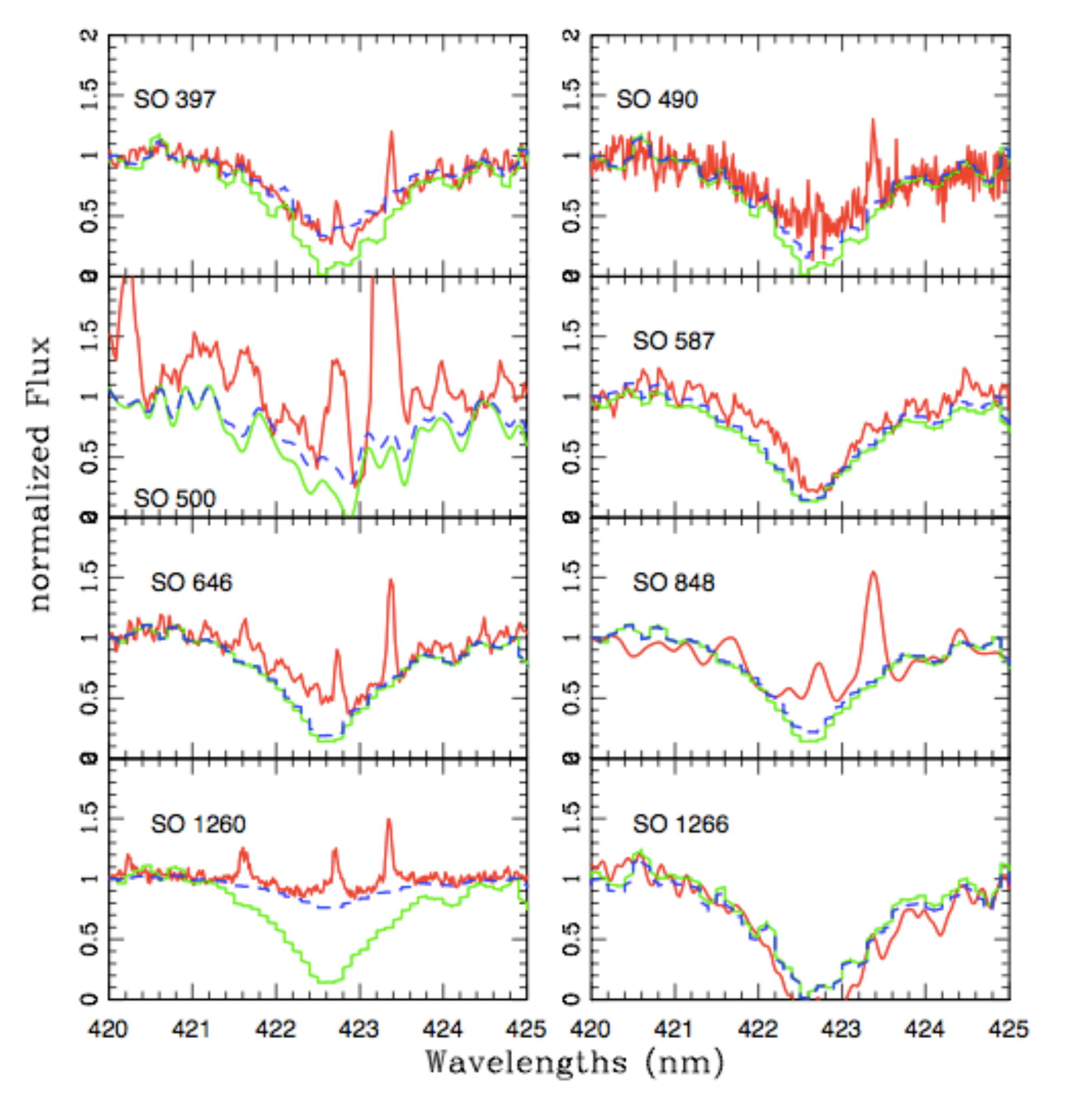}
   \caption{Ca I $\lambda$422.6 nm absorption line in the X-Shooter spectra (red-solid lines).  
The green line shows the spectrum of the adopted class III 
template and the blue-dashed line the adopted model with the emission predicted from the slab model added to the template. Fe I emission lines at 421.6, 422.7 and 423.3 nm can be also identified.}
              \label{CaI_line}%
    \end{figure}

\begin{figure}
   \centering
  \includegraphics[width=9.5cm, angle=0]{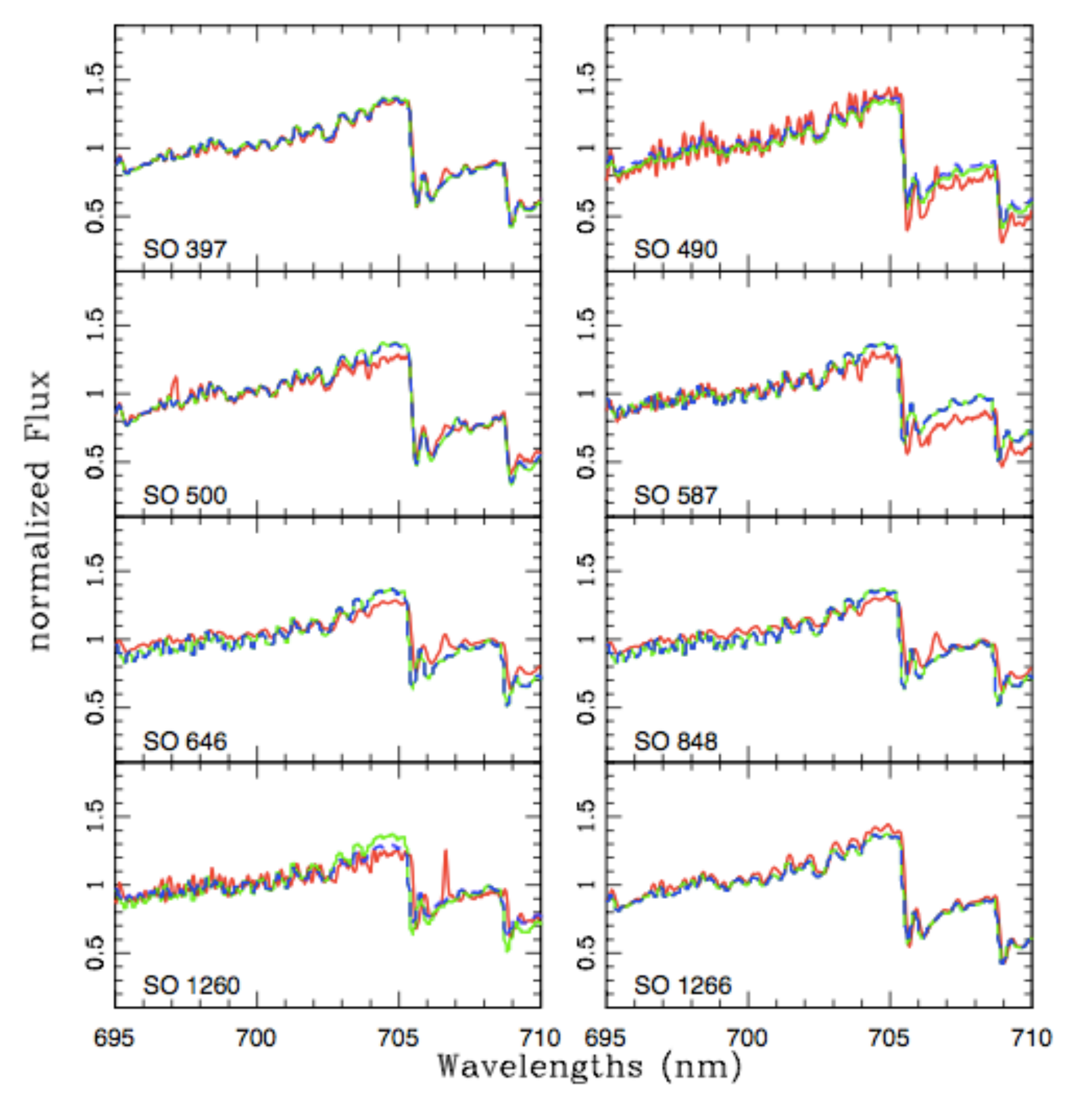}
   \caption{Region around 700 nm.  The red-solid lines shows the class II spectra, the green line shows the spectrum of the adopted class III 
template and the blue-dashed line the adopted model with the emission added to the template.}
              \label{norm700}%
    \end{figure}

The best fit models are shown in Fig.~\ref{jump_all} for the UV continuum 
and Fig.~\ref{CaI_line} for the CaI line at 422.6 nm.
Fig.~\ref{norm700} shows the normalization region around 700 nm. In 7/8 objects the excess emission is negligible; 
only in the object SO1260  this is not the case and the template normalization needs to take this into account.\\
The accretion luminosity is given by the emission of the adopted slab model scaled to fit the data as described above.  
In 6/8 objects the excess luminosity is clearly detected; in two cases (SO587 and SO1266)
we can only determine upper limits.
It is interesting to note that the Balmer continuum accounts for  about 50\% of \Laccc\ in all objects but the two strongest accretors 
(SO397, where it drops to 17\%, and SO1260, 26\%). In these two objects the excess emission is dominated by the Paschen continuum. The correction for the emission short ward of 330 nm, which is the shortest observed wavelength in most of our spectra is typically a factor of two.

The uncertainty on \Laccc\ derives firstly from the noise of the target and the template spectra, 
then from degeneracies in the slab model which affect the correction for the emission below 330 nm,   
the difficulty of constraining the emission in the Paschen continuum in objects where line veiling is not detected, 
and the choice of the class III template.
The large set of models we have computed suggests that the first is never larger than 20-30\% and the uncertainty 
introduced by the definition on the Paschen continuum is of the same order. 
Note that the slab parameters themselves are not well constrained as many trade-offs are possible, all resulting, however, in very similar
emission spectra. As far as the choice of the class III is concerned, 
we find that, when two templates provide an equally good fit to the 
observed spectra, the corresponding values of \Laccc\ are within 10\,\% of each other. 
In one case (SO848), we find that  none of our  
class\,III templates with similar spectral type reproduces the observations very well. 
However, even in this case, the two values of \Laccc\ (obtained using Sz94 and SO797 as 
templates) differ by  only 16\%.
In summary, 
a total uncertainty of 50\% is adopted for accretion luminosities \Laccc\ calculated with this method, which incorporates
uncertainties in distance, bolometric corrections and the exclusion of emission lines estimated in the 
next subsection.

\subsection* {Lines contribution to \Lacc}
\label{line_contribution}

Our estimate of \Laccc\ is
obtained from the continuum emission of the accretion slab  model  
without taking into account the contribution of the emission lines (which may vary with 
mass accretion rates and stellar parameters). 
Although the energy budget in the lines can be large, this is common practice, and 
excluding the line contribution makes our accretion luminosity estimates comparable 
to those of previous works  (HH08, Valenti et al. 1993, Calvet \& Gullbring, 1998). 

To assess the importance of line emission we have computed their contribution to the
accretion luminosity. The fraction of luminosity
in the Balmer lines with respect to the continuum accretion luminosity \Laccc\  
is reported in Table~\ref{frac_lum}, column 5.
We estimate that the high-n Balmer lines (from 12th Hydrogen recombination line 
($\lambda$=375.0 nm) to the pseudo-continuum) contain a fraction between $\sim$0.1 
and $\sim$0.4 of the continuum accretion luminosity in 5/8 of the class II objects. 
In SO500 the luminosity in the Hydrogen recombination lines is similar to the continuum 
accretion luminosity, and in SO587 and SO1266, where there is an upper limit in the 
continuum accretion luminosity, the fraction of luminosity in the Hydrogen lines is 
higher than one. This is not surprising being these lines likely due to the 
chromospheric activity in these two stars, and not to the accretion luminosity 
(see Section~\ref{notes}).  

   \begin{table*}
   \centering
      \caption[]{Accretion luminosities and mass 
      accretion rates. The first column 
      shows the name of the target, the second column the name of the template used to 
      compute the accretion luminosity. The 3rd and 4th 
      columns list the observed and the intrinsic Balmer Jump, respectively. 
In the 5th column the fraction of accretion luminosity contained in the Balmer series lines are reported, and 
${L_{high-n\,BL}}$ refers to the luminosity of the Balmer lines at wavelengths shorter than $\lambda$375.0 nm. 
      The last two 
      columns represent the accretion luminosities and the mass accretion rates 
      obtained from the excess continuum emission.}
         \label{frac_lum}
    \begin{tabular}{c c c c c c c}
    	\hline
	Name & Template & BJ$_{obs}$ & BJ$_{intr}$ & $\frac{L_{high-n\,BL}}{Lacc}$ &  log\Laccc & log\Maccc \\
			&	&		&		&		 & (\Lsun) & (\Msun/yr) \\
		\hline
SO397 &	SO797 & 0.85 & 1.9 & 0.16 	 &-2.71 &	-9.42\\
SO490 &	SO797 & 1.50 & 10.4 & 0.36 &	-3.10 &	-9.97\\
SO500 &	SO925 & 3.52& 13.1 & 1.05	 &-3.95 &	-10.27\\
SO587 &	Sz94 & 0.50 & ... &  $>$1.7	&$<$-4.00 & $<$-10.41\\
SO646 &	Sz94 & 1.23 & 14.4 & 0.20 	 &-3.00 &	-9.68 \\
SO848 &	Sz94 & 2.72 & 14.4 & 0.43 	 &-3.50 &	-10.39\\
SO1260 & Sz94 & 1.61 & 1.38 & 0.12 	 &-2.00 &	-8.97 \\
SO1266 & SO797 & 0.18 & ... & $>$1.8  &$<$-4.85 & $<$-11.38\\
		\hline
	\end{tabular}
   \end{table*}

\section{Emission lines as accretion diagnostics}
\label{emiss_section}

\begin{figure}
	\begin{center}
	
		\subfigure[\Ha\ profiles of the class II objects]{
		\label{ha2}
		\includegraphics[width=9cm]{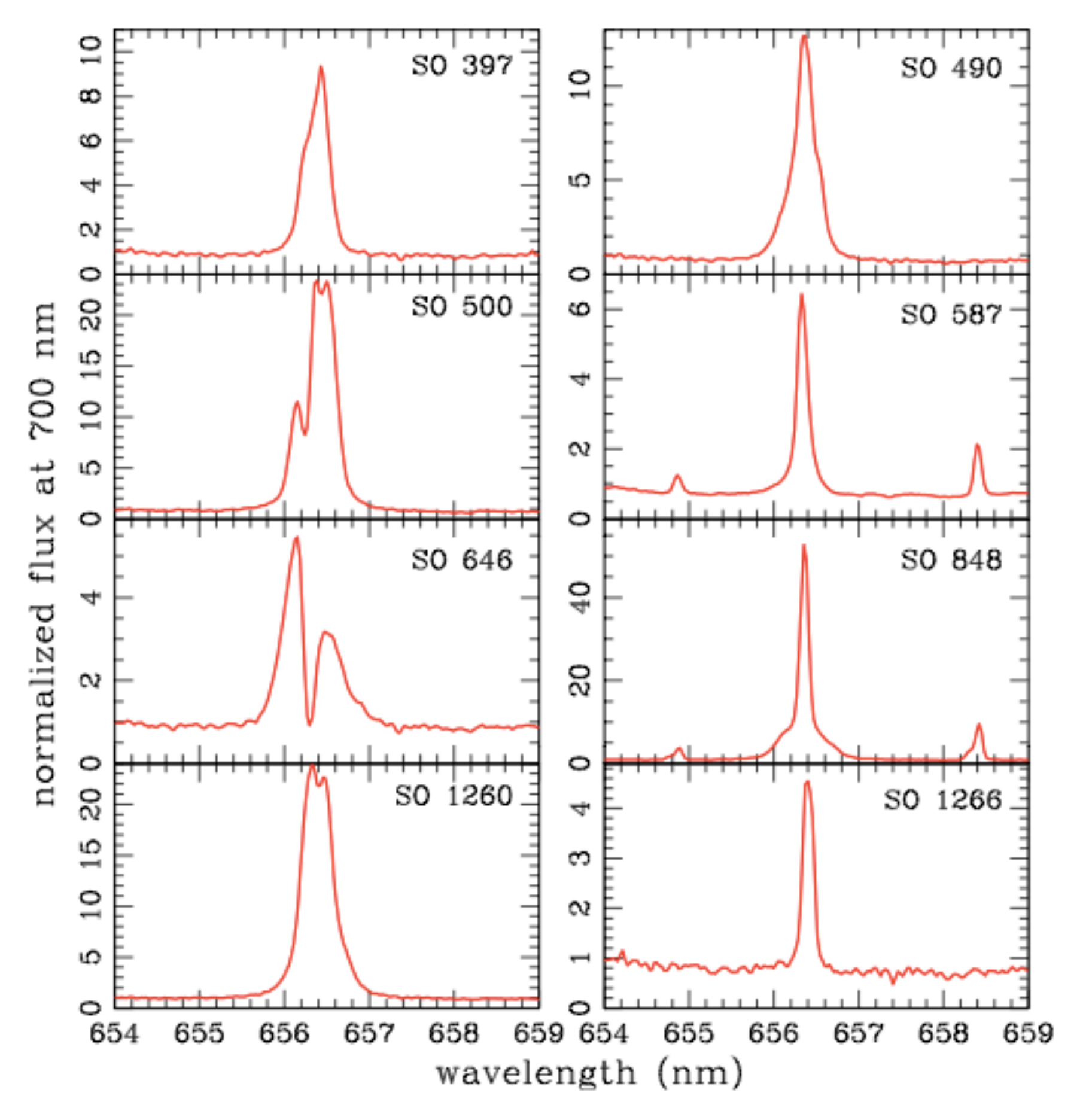}
		}

		\subfigure[\Ha\ profiles of the class III objects]{
		\label{ha3}
		\includegraphics[width=9cm]{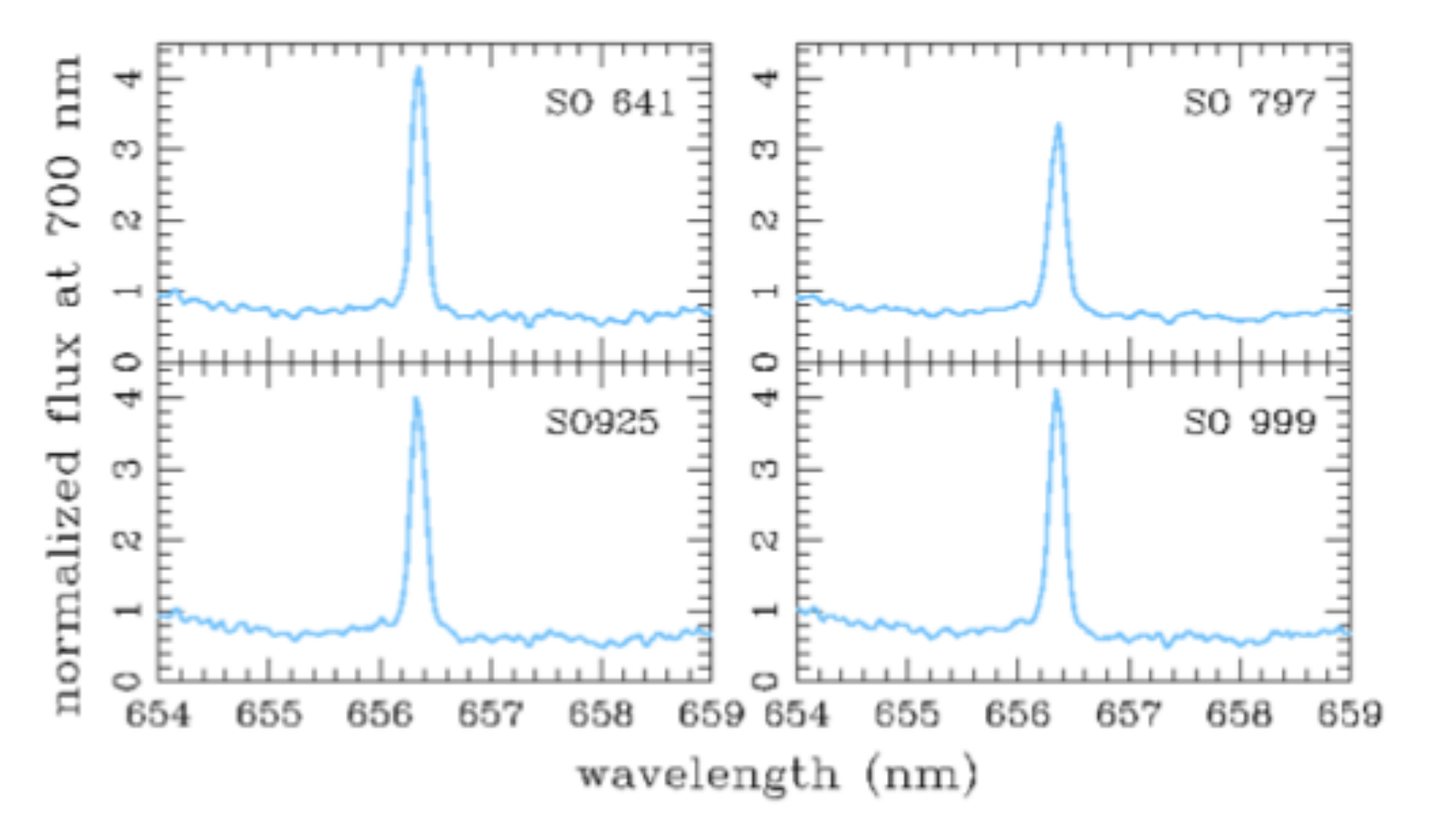}
		}
	\end{center}
	\caption{\Ha\ line profiles normalized at 700 nm (class II red lines, panel (a), and class III cyan lines, panel (b)). 
The class II line profiles exhibit diverse morphologies that presumably arise from differing accretion rates value, 
gas temperature and geometries (inclinations and magnetospheric radii).}
              \label{ha_prof}
\end{figure}

As described in Section~\ref{introduction}, 
magnetospheric accretion produces emission lines that span from UV to IR wavelengths, 
these lines can  be used to obtain an estimate of the accretion luminosity, 
and are also signatures of the chromospheric activity of the star. \\
The \Ha\ line luminosity has been found to be strictly related to the accretion luminosity 
(e.g. Muzerolle et al. 2003, Muzerolle et al. 2005, HH08, Dahm 2008, Fang et al. 2009). 
This kind of relation has been found for other hydrogen recombination lines as well 
(H$\beta$, H$\gamma$, H11, Pa$\beta$ and Pa$\gamma$; e.g. Muzerolle et al. 1998, 
Muzerolle et al 2003, Natta et al 2004, HH08, Dahm et al 2008).  
 Unfortunately, the low signal-to-noise of our NIR spectra 
 at $\lambda > 2100$ nm prevents us from detecting the Br$\gamma$ 
 line reliably, hence we do not provide measurements for this line. \\
The CaII near-IR triplet ($\lambda\lambda$849.8, 854.2, 866.2 nm) broad emission 
components are commonly attributed to accretion shocks very 
close to the stellar surface (Batalha et al., 1996). In particular, 
this triplet  has been used to detect the presence of accretion 
in low- and intermediate-mass stars (Hillenbrand et al. 1998; Muzerolle et al. 1998; Rhode et al. 2001, Mohanty et al., 2005). \\
The HeI emission line at 587.6 nm is another feature that has been used to characterize accretion processes.  
Among pre-main-sequence stars, broadened emission line profiles of HeI transition have been linked 
with magnetospheric accretion flows, as demonstrated by Muzerolle et al 1998. \\
Another line that moderately correlates with the accretion luminosity is the NaI$\lambda$589.3 nm line 
that is supposed to be produced in the magnetospheric infall zone. 
The HeI and NaI lines are usually associated with both infall and outflows 
(Hartigan et al. 1995, Beristain et al. 2001, Edwards et al. 2006) 
but  throughout this paper they are used as tracers of the accretion. \\
The width of the \Ha\ line at 10\% of the peak has been extensively used to discriminate between accreting 
and not-accreting objects and to derive an estimate of the mass accretion rate (Natta et al., 2004, White \& Basri 2003). 
In fact, as we can clearly see from Fig.~\ref{ha_prof} many of the class II objects have a broader \Ha\ line 
profile than the class III targets. 
However, in this work we decided not to use the \Ha\ 10\% line width to measure accretion rates, because, as pointed out by Natta et al. (2004), it does not provide accurate measurements of \Macc\ for individual objects.

Tables~\ref{ew_flux_1}--\ref{ew_flux_2} list fluxes and equivalent widths  (EWs) 
for the lines described above.
We measured emission line fluxes and equivalent widths by fitting Gaussian 
profiles to the observed lines with the SPLOT/IRAF task.  
Where the line does not have a Gaussian-like profile, the line flux and EW are measured by directly 
integrating the flux along a window that includes the entire line 
(Fig.~\ref{ha_prof}  presents a compilation 
of the \Ha\ line profiles for  our class\,II and class III sources).
The errors on the EWs and fluxes have been computed using a Monte Carlo approach: we added a 
normally distributed noise to the spectrum, following the error at each wavelength, and gave as error the standard 
deviation of the EW distribution of 1000 such spectra (Pascucci et al. 2011).  
The errors on the flux determination are  larger for the fainter lines, as the 
continuum calibration around these lines is more uncertain. 
Typically we estimate errors between 1-5\% for fluxes higher than 
$1.0\times10^{-16}\,{\rm erg/cm^2/s}$, and of $\sim$10\% for fainter fluxes. 

Emission lines fluxes and equivalent widths of H$\alpha$, H$\beta$, H$\gamma$, H11, 
HeI lines are measured for all the  class~II objects in our sample. 
NaI, CaII $\lambda$854.2 and CaII $\lambda$866.2 are measured for all our targets 
except one, two and three objects respectively, while we do not have an estimate of 
the Pa$\gamma$, Pa$\beta$ for SO1266 because we completely miss the portion of the 
spectrum related to the NIR arm due to problems during the observations.

\begin{sidewaystable*}
\vspace{6cm}
\centering
\caption{   Hydrogen recombination line Fluxes (erg sec$^{-1}$ cm$^{-2}$) and Equivalent Widths (nm). }
\begin{tabular}{c | c c | c c | c c | c c | c c | c c}
\hline \hline
Target & \multicolumn{2}{c}{H$\alpha$} & \multicolumn{2}{c}{H$\beta$} & \multicolumn{2}{c}{H$\gamma$} & \multicolumn{2}{c}{H11} & \multicolumn{2}{c}{Pa$\gamma$} & \multicolumn{2}{c}{Pa$\beta$} \\
 & Flux & Ew & Flux & Ew & Flux & Ew & Flux & Ew & Flux & Ew & Flux & Ew \\
\hline
SO397 & 4.25e-14 & -3.39$\pm$0.04 & 7.74e-15 & -2.33$\pm$0.03 & 4.78e-15 & -2.07$\pm$0.05 & 1.25e-15 &	-1.29$\pm$0.11 & 3.64e-15 & -0.09$\pm$0.03 & 2.08e-15 & -0.06$\pm$0.04 \\
SO490 & 1.56e-14 & -5.45$\pm$0.06 & 2.92e-15 & -3.86$\pm$0.02 & 1.99e-15 & -4.11$\pm$0.04 & 6.11e-16 &	-1.98$\pm$0.21 & 1.16e-15 &	-0.073$\pm$0.03 & 4.55e-16 & -0.031$\pm$0.03 \\ 
SO500 & 7.47e-15 & -12.15$\pm$0.15 & 8.71e-16 & -8.09$\pm$0.04 & 3.53e-16 & -6.26$\pm$0.13 & 1.22e-16 & -3.20$\pm$0.29 & 3.61e-16 & -0.09$\pm$0.09 & 5.19e-16 & -0.14$\pm$0.07 \\
SO587 & 2.84e-14 & -1.47$\pm$0.11 & 5.98e-15 & -1.04$\pm$0.03 & 2.88e-15 & -0.69$\pm$0.05 & 6.56e-16 &	-0.42$\pm$0.07 & 2.9e-16	& -0.004$\pm$0.17 & 4.6e-16 & -0.007$\pm$0.04 \\
SO646 & 2.32e-14 & -2.34$\pm$0.16 & 3.54e-15 & -1.01$\pm$0.02 & 2.67e-15 & -1.27$\pm$0.04 & 1.43e-15 & -1.28$\pm$0.11 & 2.26e-15 & 	-0.096$\pm$0.05 & 2.87e-15& -0.14$\pm$0.03 \\
SO848 & 1.79e-14 & -8.35$\pm$0.50 & 4.03e-15 & -9.74$\pm$0.02 & 1.80e-15 & -6.09$\pm$0.08 & 3.09e-16 &	-2.01$\pm$0.21 & 5.01e-16  & -0.09$\pm$0.05 & 6.91e-16 & -0.12$\pm$0.04 \\
SO1260 & 1.43e-13 & -11.71$\pm$0.08 & 2.58e-14 &	-4.43$\pm$0.02 & 1.62e-14 & -3.42$\pm$0.02 & 4.37e-15 & -0.94$\pm$0.02 & 9.39e-15 & -0.34$\pm$0.04 & 1.33e-14 & -0.52$\pm$0.04 \\
SO1266 & 3.96e-15 & -0.84$\pm$0.04 & 9.37e-16 & -0.97$\pm$0.03 & 4.37e-16 & -0.68$\pm$0.06 & 6.74e-17 & -0.51$\pm$0.03 & -- & -- & -- & -- \\
\hline \hline
\end{tabular}
\label{ew_flux_1}
\end{sidewaystable*}

\begin{table*}
\centering
\caption{ Fluxes  (erg sec$^{-1}$ cm$^{-2}$) Ew (nm) for accretion indicators. 
}
\begin{tabular}{c | c c | c c | c c | c c }
\hline \hline
Target & \multicolumn{2}{c}{HeI$_{\lambda 587.6}$} & \multicolumn{2}{c}{NaI$_{\lambda 589.3}$} & \multicolumn{2}{c}{CaII$_{\lambda 854.2}$} & \multicolumn{2}{c}{CaII$_{\lambda 866.2}$}  \\
 & Flux & Ew & Flux & Ew & Flux & Ew & Flux & Ew \\
\hline
SO397 & 1.17e-15 & -0.23$\pm$0.01 & 1.61e-16 & -0.05$\pm$0.03 & 1.124e-15 & -0.046$\pm$0.02 & 1.36e-15 & -0.02$\pm$0.03 \\
SO490 & 4.13e-16 & -0.47$\pm$0.01 & 2.87e-17 & -0.045$\pm$0.02 & 1.55e-16 & -0.018$\pm$0.02 & $<$8.0e-17 & $<$-0.008 \\
SO500 & 3.65e-17 & -0.21$\pm$0.03 & 2.55e-17 & -0.19$\pm$0.07 & 5.05e-16 & -0.29$\pm$0.03 & 4.002e-16 & -0.19$\pm$0.02 \\
SO587 & 5.84e-16 & -0.062$\pm$0.01 & 2.13e-16 & -0.034$\pm$0.03 & $<$3.0e-16 & $<$-0.003 & $<$3.0e-16 & $<$-0.003 \\ 
SO646 & 1.14e-15 & -0.20$\pm$0.01 & $<$4.3e-17 & $<$0.007 & 1.12e-15 & -0.065$\pm$0.01 & 7.20e-16 & -0.039$\pm$0.02 \\
SO848 & 3.93e-16 & -0.47$\pm$0.03 & 2.61e-17 & -0.034$\pm$0.04 & 1.79e-15 & -0.45$\pm$0.01 & 1.43e-15 & -0.33$\pm$0.02 \\
SO1260 & 2.25e-15 & -0.31$\pm$0.02 & 6.37e-16 & -0.09$\pm$0.02 & 1.08e-14 & -0.51$\pm$0.01 &  7.88e-15 & -0.34$\pm$0.02 \\
SO1266 & 1.067e-16 & -0.075$\pm$0.02 & 3.81e-17 & -0.078$\pm$0.06 & $<$1.4e-16 & $<$-0.013 & $<$1.4e-16 & $<$-0.013 \\
\hline \hline
\end{tabular}
\label{ew_flux_2}
\end{table*}

Once measured the emission line fluxes $F_{line}$ from the flux-calibrated spectrum, 
the line luminosities are given by: 
\begin{equation}
{L_{line}} = 4\pi D^{2} {F_{line}} 
\end{equation}
where $D$ is  again the adopted distance of $360$\,pc. 
To determine accretion luminosities using the measured line fluxes, we use  
empirical relationships between the observed line luminosities and 
the accretion luminosity or mass accretion rates (HH08, 
Mohanty et al.~2005, Fang et al.~2009, Natta et al.~2004, Gullbring et al.~1998).  
These relationships were calibrated on relatively small samples of well-studied  
 T Tauri stars and Brown Dwarfs, for which the accretion luminosity was derived by fitting the 
observed optical veiling with  magnetospheric accretion models (Gullbring et al.~1998 for 
low-mass stars in the mass range between $\sim$0.25--0.9 \Msun, and
Calvet et al.~2004 for intermediate mass T Tauri stars in the mass range between 
1.5--4\Msun), or, for brown dwarfs, the \Ha\ line profile (Muzerolle et al.~2003).  

 According to these studies, 
the accretion luminosity can be computed from the line luminosity as 
\be
log({L}_{acc}/L_{\odot}) = b + a\times log(L_{line}/L_{\odot})
\ee
where the coefficients $a$ and $b$  are summarized in Table~\ref{ab_coeff}
 together with their origin in the literature. 
In the case of the CaII lines the equivalent widths were converted into line flux per 
stellar surface area, and then correlated directly with \Macc: 
\be
log{\dot{M}_{\rm acc}} = d + c\times log(F_{CaIIline})
\ee

\begin{table}
\centering
\caption{Line luminosity vs. accretion luminosity relationships 
for selected accretion indicators. The first column shows the name of the indicator and the corresponding wavelength in nm. 
The U-band relationship is based on a correlation between the accretion luminosity and the excess luminosity in the U photometric band.
$^1$Note that the line H11 is erroneously defined as H9 in HH08. 
}
\begin{tabular}{c c c c}
\hline \hline
 Line & $a$ & $b$ & Reference \\
\hline
H$\alpha$ (656.3) & 1.25 $\pm$ 0.07 & 2.27 $\pm$ 0.23 & Fang et al.09\\
H$\beta$ (486.1) & 1.28 $\pm$ 0.05 & 3.01 $\pm$ 0.23 & Fang et al. 09\\
H$\gamma$ (434.1) & 1.24 $\pm$ 0.04 & 3.0 $\pm$ 0.2 & HH08\\
H$11$ (377.1)$^1$ & 1.17 $\pm$ 0.03 & 3.4 $\pm$ 0.2 & HH08 \\
HeI (587.6) & 1.42 $\pm$ 0.08 & 5.20 $\pm$ 0.38 & Fang et al. 09 \\
Pa$\beta$ (1280) & 1.36 $\pm$ 0.20 & 4.00 $\pm$ 0.20 & Natta et al. 04\\
Pa$\gamma$ (1090) & 1.36 $\pm$ 0.20 & 4.10 $\pm$ 0.20 & Gatti et al. 08\\
NaI (589.3) & 1.09 $\pm$ 0.11 & 3.3 $\pm$ 0.7 & HH08 \\
Uband & 1.09$^{+0.04}_{-0.18}$ & 0.98$^{+0.02}_{-0.07}$ & Gullbring et al. 98 \\ 
\hline 
\hline
 Line & $c$ & $d$ & Reference \\
\hline
CaII (854.2) & 1.28 & -16.6 & HH08 \\
CaII (866.2) & 0.71 & -12.66 & Mohanty et al. 05 \\
\hline \hline
\end{tabular}
\label{ab_coeff}
\end{table}

Table~\ref{accrLum} reports the calculated accretion luminosity for each star based 
on the different emission lines.
 To distinguish these values from those obtained from the continuum slab model,
we denote the line-based accretion luminosities with \Laccl\ throughout the remainder
of the paper.
 Table~\ref{accrLum} lists also the  weighted  
mean accretion luminosity obtained from the secondary indicators, 
$\langle$\Laccl $\rangle$. 
\begin{table*}
\vspace{5cm}
\centering
\caption{  Estimated log\Lacc\ (\Lsun) from the secondary accretion indicators. The first column shows the star name.    
The last column lists the weighted mean log\Lacc\ based on all the accretion luminosity obtained from the secondary indicators and corresponds to the green lines in Fig.~\ref{Lacc_tot_average}.
The other columns show log\Lacc\ obtained from the emission lines, as described in the text, and correspond to the red points in  Fig.~\ref{Lacc_tot_average}.  }
\label{accrLum}
\begin{tabular}{c c c c c c c c c c c c}
\hline \hline
 Source &  Log\Lacc  & Log\Lacc  & Log\Lacc  & Log\Lacc  & Log\Lacc  & Log\Lacc  & Log\Lacc  & Log\Lacc  & Log\Lacc  & Log\Lacc & $\langle$Log\Laccl$\rangle$ \\
            &  H$\alpha$  & H$\beta$  & $H\gamma$ &  H11 & HeI & Pa$\gamma$ & Pa$\beta$ & NaI & CaII 854 & CaII 866 & \\
\hline
SO397 &  	-2.44 & -2.76 & -2.85 &	-2.80 & -2.37 & -2.48 & -2.91 & -3.45 &  -3.84 & -2.54 & -2.83 $\pm$ 0.29	\\
SO490 & 	-2.98 & -3.31	& -3.32 &	 -3.17 & -3.01 & -3.15 & -3.81 & -4.26 & -4.84 & $<$-3.4 & -3.43 $\pm$ 0.37	\\
SO500 & 	-3.39 & -3.98	& -4.25 &	 -3.98 & -4.51 & -3.84 & -3.73 & -4.32 & -3.98 & -2.92 & -3.83 $\pm$ 0.27	\\
SO587 & 	-2.66 & -2.91	& -3.12 &	 -3.13 & -2.80 & -3.97 & -3.80 & -3.32 & $<$-5.3 & $<$-3.5 & -3.15 $\pm$ 0.39	\\
SO646 & 	-2.77 & -3.19	& -3.17 &	 -2.73 & -2.39 & -2.76 &  -2.72 &	$<$-4.07 & -3.74 & -2.69 & -2.80 $\pm$ 0.23\\
SO848 & 	-2.91 & -3.12	& -3.38 &	 -3.51 & -3.04 & -3.65 & -3.56 &	4.31 & -2.36 & -1.76 & -3.13 $\pm$ 0.22\\
SO1260 & -1.78 &	-2.09 & -2.19 & -2.17 &	-1.96 & -1.92 & -1.81 &	2.80 & -1.98 & -1.55 & -1.99 $\pm$ 0.15\\
SO1266  &	-3.73 & -3.94 & -4.14 &	 -4.29 & -3.85 & -- & -- & -4.13 & $<$-4.9 & $<$-3.3 & -3.91 $\pm$ 0.23\\
\hline \hline
\end{tabular}
\end{table*}

 The accretion luminosity can be converted into mass accretion rate, \Macc\, 
 according to  
\begin{equation}
{\dot M_{acc}} = \big(1 - {{R_{*}} \over {R_{in}}}\big)^{-1}  {{L_{acc} \, R_{*}}\over{G \, M_{*}}} \sim 1.25 {{L_{acc} \, R_{*}}\over{G \, M_{*}}} ,
\label{Macc_equation}
\end{equation}
where G is the universal gravitational constant and 
the factor $(1-{{R_{*}} \over {R_{in}}})^{-1}$ $\sim$ 1.25 is estimated by assuming 
that the  accreting gas falls onto the star from the truncation radius of the disk 
($R_{in}$ $\sim$5$R_{*}$; Gullbring et al.~1998).  
The error introduced by this assumption on the measured mass accretion rates, 
considering that $R_{in}$ for a pre-main sequence star can span from 3 to 8 $R_{*}$,  is less than 20\%.
The stellar masses and radii are listed in Table~\ref{table_target_properties}. 
Note that the values of \Lacc\ for the two CaII IR lines are derived from the mass accretion rate, computed from Eq. (3), and the standard relation between \Lacc\ and\ Macc\ (Eq. 4). 
\\

\subsection{Comparison of the accretion indicators}
\label{sect_comparison}

   \begin{figure*}
   \centering
   \includegraphics[width=12cm]{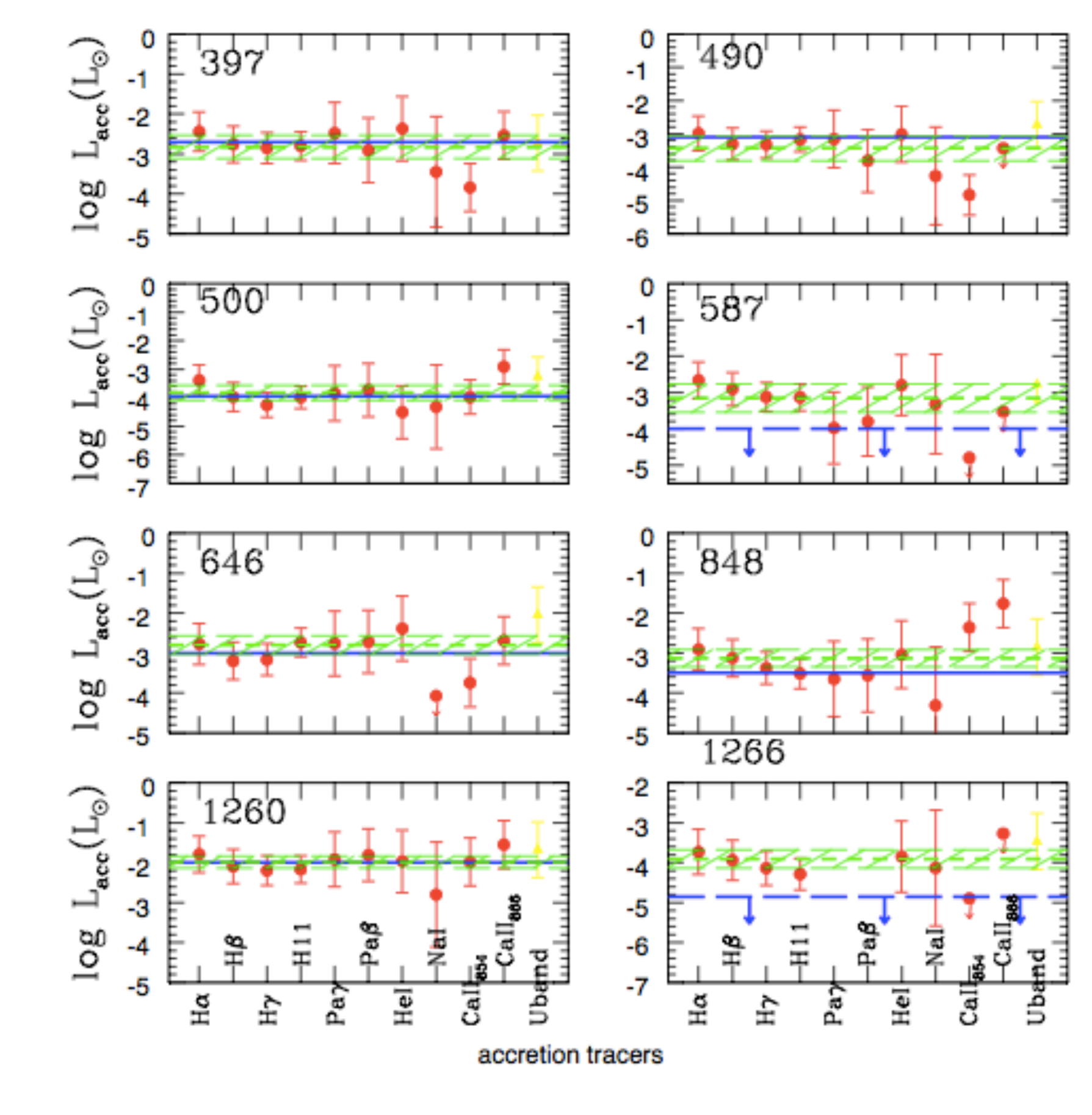}
   \caption{Comparison between all the accretion indicators listed in 
   Tab.~\ref{ab_coeff}. 
The red dots are our estimates of the accretion luminosity, 
   the red arrows are the upper limits, the yellow triangles refer to \Lacc\ obtained from 
   the U-band excess emission, which  was observed non-simultaneously with the 
   other indicators. The green dashed line is the average \Laccl\ computed considering 
   all the secondary indicators; the green shaded region is the 1$\sigma$ uncertainty 
   in the average \Laccl. The blue  line is \Laccc\ obtained using the excess continuum emission  
   (dashed blue line for the upper limits). }
              \label{Lacc_tot_average}%
    \end{figure*}

Fig.~\ref{Lacc_tot_average}  displays the \Laccl\ values of Table~\ref{accrLum}
(red circles) and also the accretion luminosity derived by Rigliaco et al. (2011) using the non-simultaneous 
$U$-band excess emission  (yellow triangle).  
Even  though all accretion indicators  except for the $U$-band have been 
observed simultaneously, for a given star the spread 
between different values of \Laccl\ is quite large,  typically  
about one order of magnitude. However, as we  have shown 
for the brown dwarf SO500 in Rigliaco et al.~(2011b), for most of the stars analyzed here 
all values are consistent within the large error bars, and the average 
\Laccl\ has a smaller uncertainty (green line in Fig.~\ref{Lacc_tot_average}). 
In particular, the total errors on  \Laccl\ for each indicator are due to the error on the line fluxes, 
on the adopted distance and on the relationship used to estimate \Laccl. 
The uncertainties on the flux and distance (which are about a factor 1.5 on the total accretion luminosity) 
are negligible with respect to the error due 
to the assumed relationship, which corresponds to an uncertainty on \Laccl\ of about a factor three. 
The uncertainty on $\langle$\Laccl $\rangle$ (which excludes by our definition  
the estimate from the U-band excess emission) computed averaging the accretion luminosity for each indicator 
weighted by the corresponding error and neglecting the upper limits, is about a factor two. 

In five stars of the sample discussed in this paper,  SO397, SO490, 
SO500, SO646 and SO1260, the accretion luminosity obtained from the Balmer 
excess continuum (blue line) 
is within the 1$\sigma$ uncertainty of the average accretion luminosity obtained by using the secondary accretion indicators. 
The three cases for which we do not find  good 
agreement will be discussed in Sect.~\ref{notes}.

  \begin{figure}
   \centering
   \includegraphics[width=8cm]{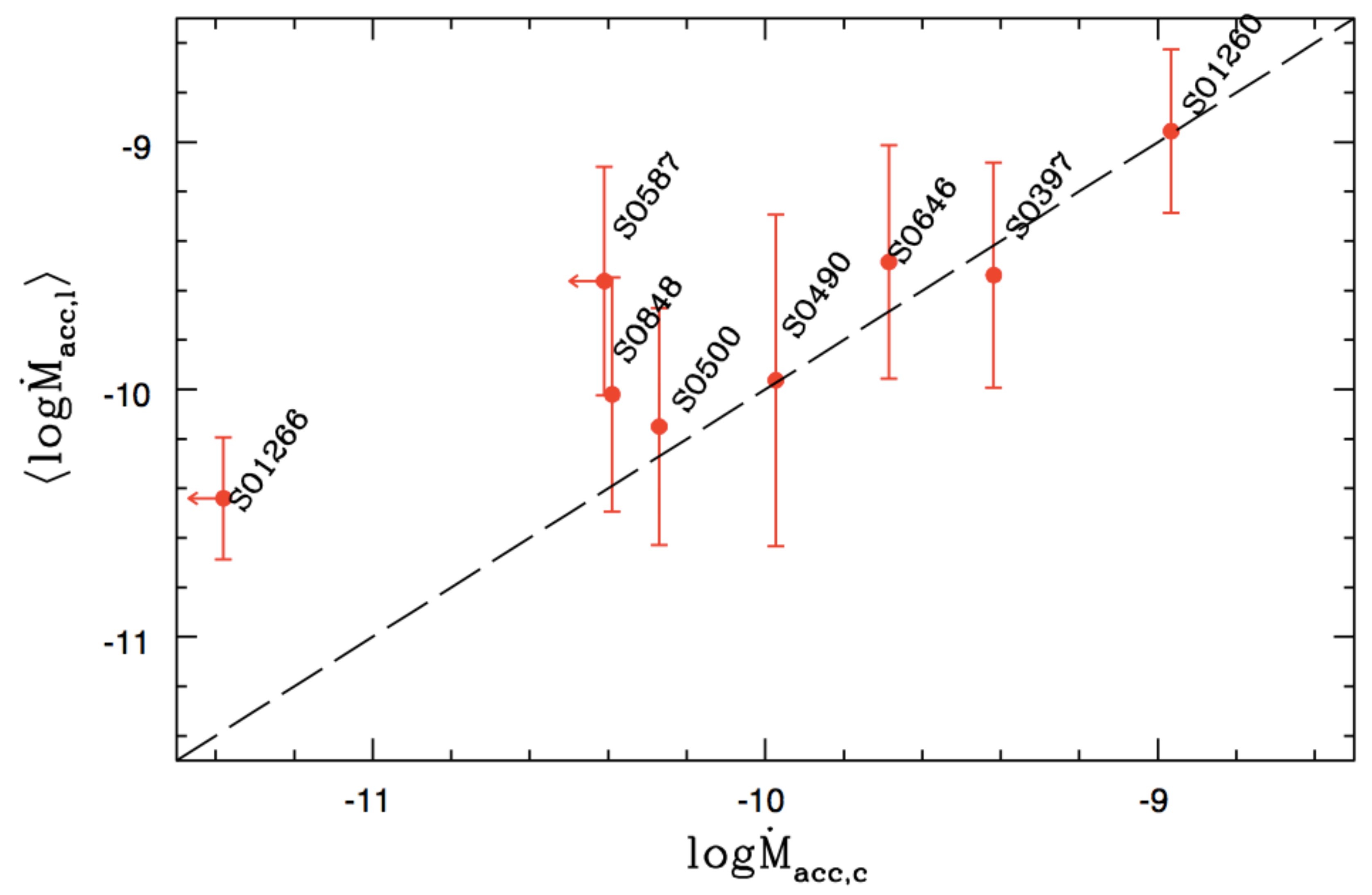}
   \caption{Comparison between the mass accretion rates computed from the average of all the indicators (\Maccl) and \Maccc.}
            \label{Macc_comparison}
    \end{figure}

In Figure~\ref{Macc_comparison} we show  for each star the 
average  mass accretion rate derived from \Laccl\ with Eq.4 (\Maccl)  
versus the mass accretion rate obtained with the same formula from 
the continuum accretion luminosity (\Maccc).  
From this small sample we can conclude that the two quantities are in agreement 
within the 1$\sigma$ error (except for the cases discussed in Sect.~\ref{notes})
and do not show any robust trend with increasing \Maccl.  
However, the sample considered here is small, and this issue 
could be explored further on a bigger sample of objects.

\subsection{Reliability of the accretion tracers}
\label{reliab}

One way to investigate the reliability of the accretion tracers, and the relation between 
accretion continuum luminosity (\Laccc) and the accretion luminosity from the 
secondary indicators (\Laccl) is comparing \Laccc\ with the secondary tracers line luminosities or fluxes, 
including the sample analyzed here 
with those available in  the literature.  
We have recomputed the relationships between \Laccl\ and \Laccc\ after adding our
new X-Shooter data for $\sigma$\,Ori to the historical samples of accreting TTS
published previously. These results are given in Table~\ref{new_coeff} and are
discussed in the following subsections.

\subsubsection{H$\alpha$, H$\beta$ and HeI}
\label{Ha_Hb}

\begin{figure}	
   \centering
   \includegraphics[width=9.5cm]{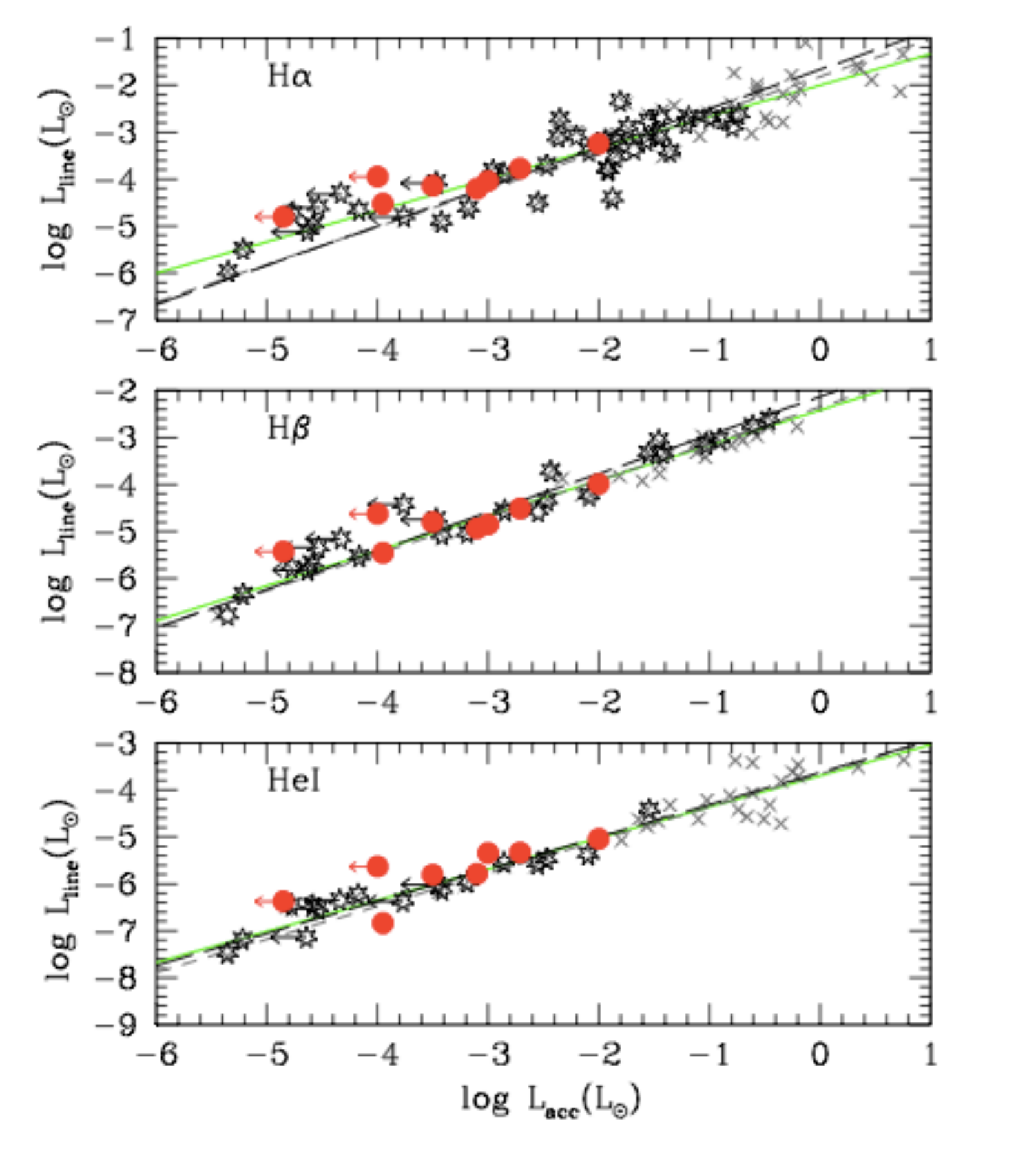}
   \caption{  Relationship between selected line luminosities and the accretion continuum luminosity. 
The red dots and arrows are the sample analyzed in this paper,  the black asterisks and arrows refer to the sample analyzed by HH08. The grey crosses are the sample collected by Fang et al. (2009).  The grey short-dashed line corresponds to the relation found by Fang et al. (2009). The black long-dashed line is the relation found by HH08, computed ignoring the upper limits in \Lacc. The green lines correspond to the relations found using the actual detection of the sample analyzed here combined with the Fang et al. (2009) sample.  }
              \label{HaHbHei}%
    \end{figure}

 For the \Ha, H$\beta$ and HeI lines as accretion tracers,  
we refer to Fang et al. (2009). They collect young stellar objects with 
measured \Ha, H$\beta$ and HeI emission line luminosities from literature 
(Gullbring et al. 1998, HH08 and Dahm et al. 2008 for \Ha, Gullbring et al. 1998 and 
HH08 for H$\beta$, and HH08 and Dahm et al 2008 for HeI) for 
a sample of low-mass stars and brown dwarfs in the Taurus Molecular Cloud and 
 the young open cluster IC\,348.
Fig.~\ref{HaHbHei} shows the comparison between the relationships found by Fang et al. (2009) 
(grey short-dashed line, and grey crosses), 
 by HH08 (black long-dashed line and black asterisks), and 
 by us (green solid line and red dots). 
The correlation found by HH08 and  by  
Fang et al. (2009) have been computed by least square fitting the  
point distributions and neglecting the upper limits. 
We did the same  adding to the Fang et al. (2009) sample the targets analyzed in 
this paper. 

\begin{table}
\centering
\caption{ Newly determined line luminosity vs accretion luminosity relationships 
for selected accretion indicators. 
}
\begin{tabular}{c c c }
\hline \hline
 Line & $a$ & $b$ \\
      &     &     \\
\hline
H$\alpha$ & 1.49 $\pm$ 0.05 & 2.99 $\pm$ 0.16 \\
H$\beta$ & 1.34 $\pm$ 0.02 & 3.01 $\pm$ 0.09 \\
H$\gamma$ & 1.30 $\pm$ 0.02 & 3.32 $\pm$ 0.08 \\
H$11$ & 1.25 $\pm$ 0.03 & 3.75 $\pm$ 0.13  \\
HeI & 1.51 $\pm$ 0.03 & 5.59 $\pm$ 0.17 \\
Pa$\beta$ & 1.49 $\pm$ 0.04 & 4.59 $\pm$ 0.14 \\
NaI & 0.76 $\pm$ 0.10 & 3.76 $\pm$ 0.47  \\
Uband & 1.33 $\pm$ 0.06 & 1.55 $\pm$ 0.18  \\ 
\hline 
\hline
 Line & $c$ & $d$  \\
      &     &     \\
\hline
CaII854 & 1.90 $\pm$ 0.06 & -20.37 $\pm$ 0.78 \\
CaII 866 & 1.41 $\pm$ 0.08 & -17.76  $\pm$ 0.89 \\
\hline \hline
\end{tabular}
\label{new_coeff}
\end{table}

From Fig.~\ref{HaHbHei} it can be seen 
that there is a very good agreement  
between the empirical calibration for the H$\beta$ and HeI lines 
presented in  the literature  and the new one obtained including our targets,
while 
the relation log\Laccc\ -- log$L_{\rm H\alpha}$ found including our objects is flatter 
toward  the low-accretion luminosity regime with respect to those found in the 
past.  
The reason for this flattening could lie in the active chromosphere of class II objects (Houdebine 
et al. 1996, Franchini et al. 1998, Ingleby et al. 2011b, see Fig.~\ref{chrom_act}). 
The chromospheric contribution to the measured accretion luminosity (from excess 
continuum or from emission lines) is expected to be small and negligible when compared 
to that due to the accretion 
onto the class II stars. However, as the accretion luminosity and mass accretion rate decrease, the 
contribution of the chromosphere, which stays approximately constant in the age range 
1-10 Myr (Ingleby et al. 2011a), becomes comparable or dominant with respect to the accretion. 
We briefly speculate on this issue using 
the \Ha\ line for our sample. 

From the flux-calibrated spectra of the four class III stars in our sample, we have 
measured the flux of the \Ha\ line  and using the relation reported in 
Table~\ref{ab_coeff} we have estimated the corresponding \Lacc\ we should expect 
if the line were produced in the  magnetospheric accretion framework rather than in  
the stellar chromosphere. The average value for the class\,III stars is 
log\Lacc$_{,H\alpha,chrom}$=-4.1. 
We expect that the 
 chromospheric emission becomes dominant already for somewhat higher values 
when the chromospheric contribution starts to be 
between the 10\% and the 100\% of the accretion luminosity. 
We do not see clear evidence of  such a flattening for the other emission lines. 
 This leads us to speculate that as
the accretion luminosity becomes lower and lower, the first accretion tracer 
which starts to suffer from 
 chromospheric contamination is the \Ha\ line, as expected for a typical chromospheric spectrum. 
A detailed analysis of the class III activity from the targets observed with X-Shooter during the GTO 
survey of star forming regions is being done in Manara et al. (2012, in prep).

\subsubsection{Other Hydrogen recombination lines}

\begin{figure}	
   \centering
   \includegraphics[width=9.5cm]{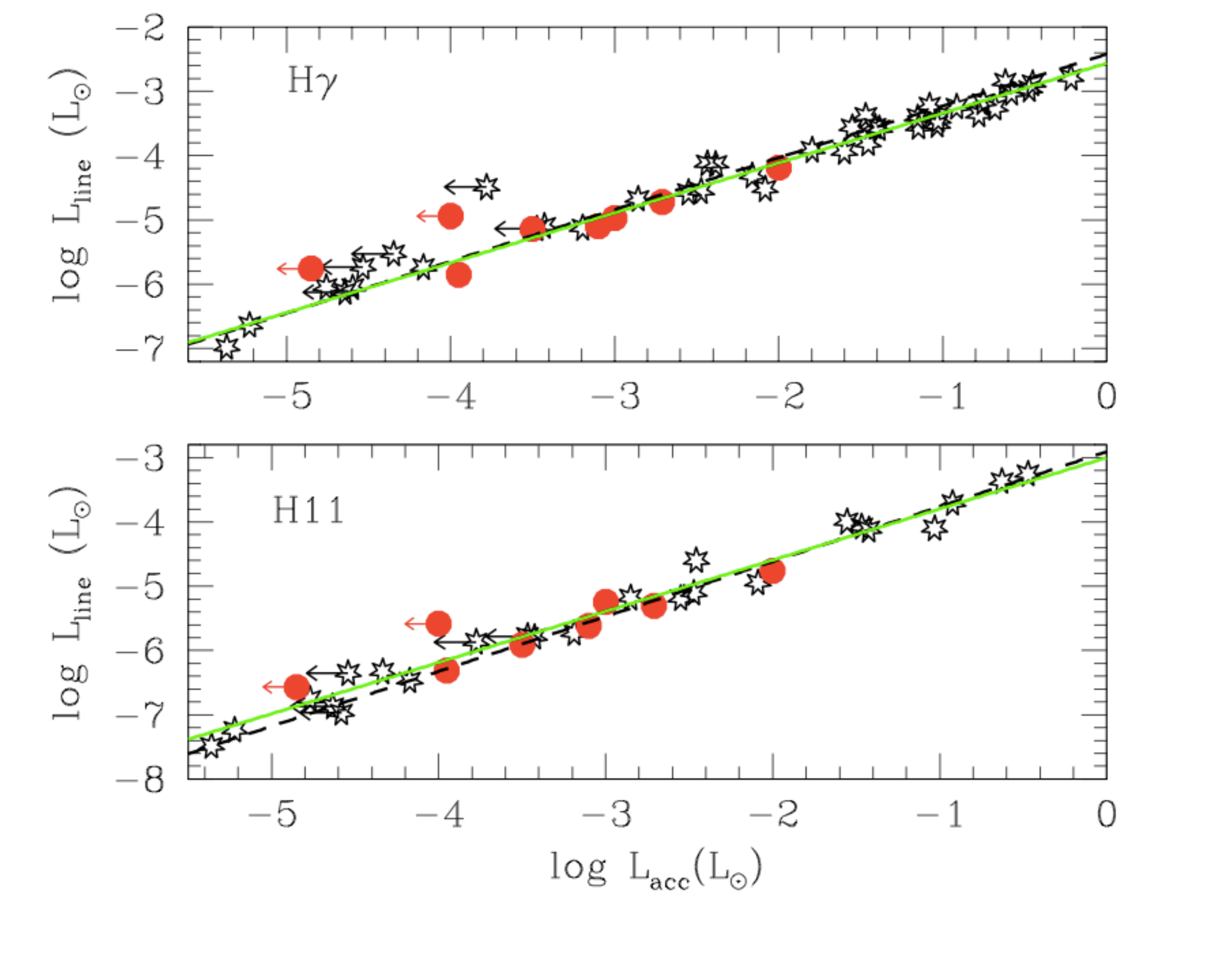}
   \caption{ Relationship between selected line luminosities and the accretion continuum luminosity. 
The red dots and arrows are the sample analyzed in this paper,  the black asterisks refer to the sample analyzed by HH08, where they found the relation drawn as black line. The green solid line corresponds to the relation found using also the actual detection of our sample. }
              \label{comp_hg_h9}%
    \end{figure}

We also recomputed the relationships 
 between \Laccc\ and the logarithm of $L_{\rm H\gamma}$ and $L_{\rm H11}$,
respectively, that were previously discussed by HH08.
We fitted our points with those collected by HH08 ignoring the upper limits. 
The  data and the old and new relationships 
are shown in Fig.~\ref{comp_hg_h9}.  

\begin{figure}
   \centering
   \includegraphics[width=9.0cm]{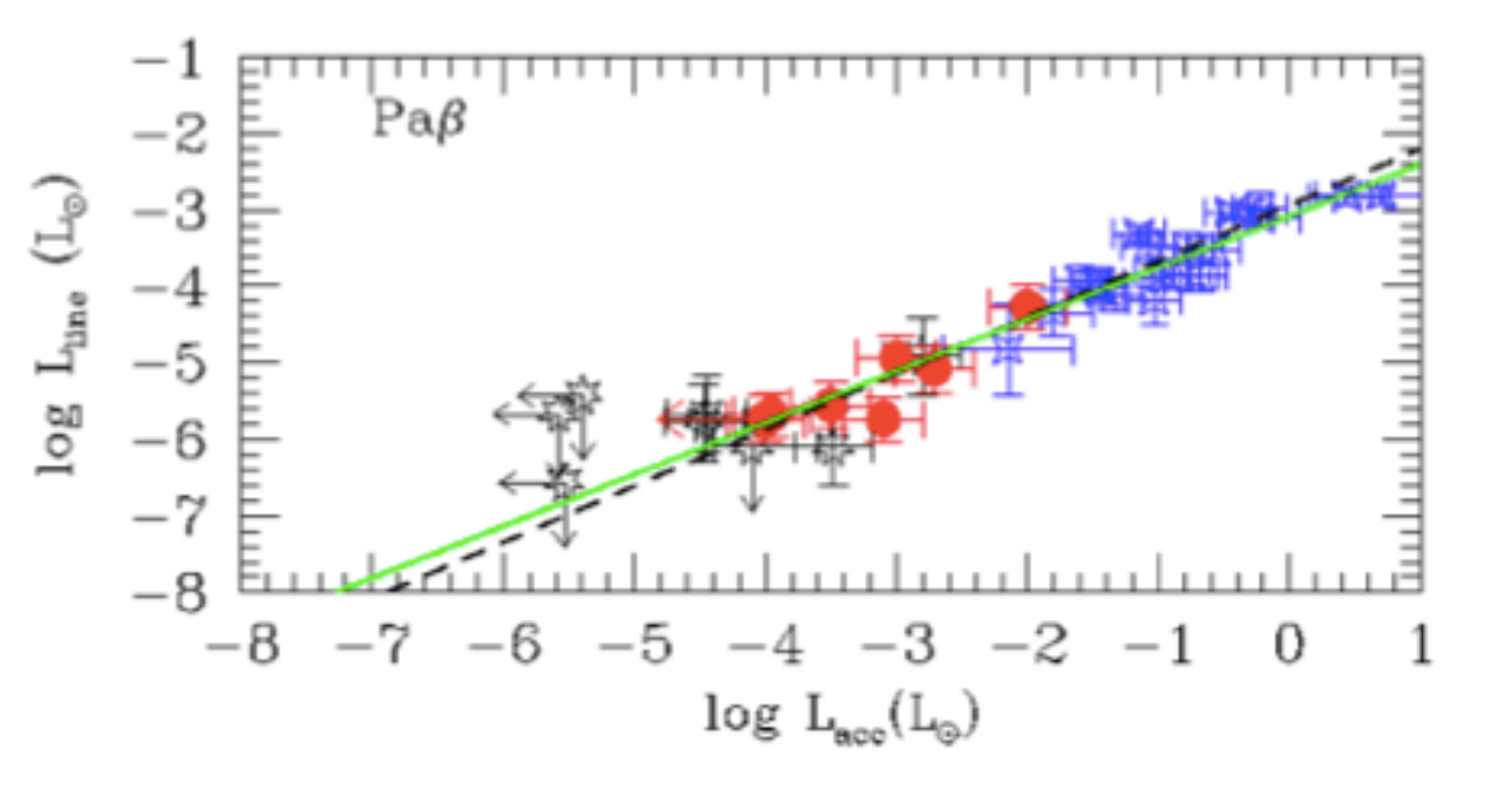}
   \caption{\Pab\ luminosity as a function of \Laccc. The black asterisks  are the sample from Natta et al.~(2004), the blue diamonds 
represent the sample analyzed by Muzerolle et al. (1998). The solid black line is the best-fitting relation found by Natta et al.~(2004) (Eq. (2) of their paper). The green triangles are the objects of our sample, and the green line is the relation found combining these objects with the others. }
              \label{pab_comp}%
    \end{figure}

In Fig.~\ref{pab_comp} we show the correlation between \Laccc\ and the \Pab\ line 
luminosity for the targets discussed in this paper, and for previous samples 
 from the Ophiuchus and Cha\,I star forming regions analyzed 
by Natta et al.~(2004) and Muzerolle et al.~(1998). 
The trend found by Natta et al.~(2004) shows a good correlation over the whole range of masses from few tens of Jupiter masses to about one solar mass. 
Our new data confirm the trend found by Natta et al. (2004) and reinforce the strength of this line as an accretion indicator, 
agreeing with the results found by Antoniucci et al. (2011) who also tested the reliability of this indicator as accretion tracer.

\subsubsection{Other accretion tracers}

\begin{figure}
   \centering
  \includegraphics[width=9.0cm]{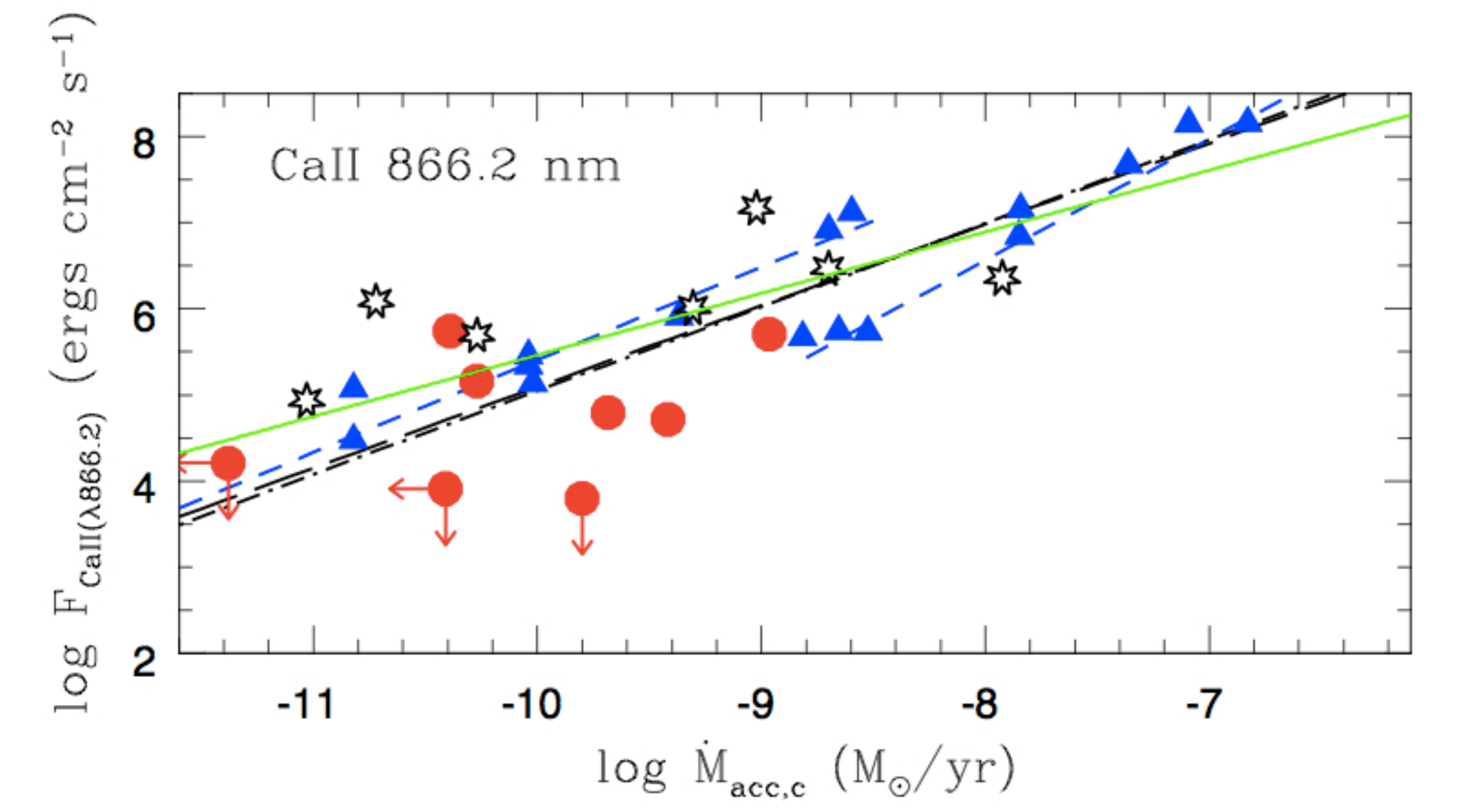}
   \caption{\Maccc\ versus flux of the CaII $\lambda$866.2 line. The blue triangles and the blue lines (long- and short-dashed) represent the data set and the correlation found by Mohanty et al. (2005). 
The black asterisk are the sample taken into account by HH08 and the blue dotted-dashed lines are the relationships found in their paper. The red points are the objects analyzed in this paper. The green solid line 
is the least square fit of all the targets plotted in the figure, neglecting the upper limits.}
              \label{caii_comp}%
    \end{figure}

A quite large spread can be seen in Fig.~\ref{caii_comp} for the mass accretion rates retrieved using the CaII 866 nm line. 
We have compared  our results with those obtained by Mohanty et al.~(2005) and HH08. 
Fig.~\ref{caii_comp} shows their relationships. In particular, the black 
long-dashed line represents the linear fit of the data collected  by
Mohanty et al. (2005).  These authors, moreover, suggest  
that there might be a small systematic offset between the higher-mass accretors and the 
 very low-mass sample, with the latter possibly showing a somewhat higher CaII emission for a given 
\Macc.  Consequently, they performed separately fits for the higher- and lower-mass samples. 
Mohanty et al. (2005) speculate that the reason of this offset could lie in the 
methodology of \Macc\ derivation for the intermediate-mass objects (U-band veiling)  with respect
to the low-mass objects (\Ha\ modeling), 
and/or in  a different formation region of the CaII line in the two mass regimes 
(infalling gas or accretion shock). 

We have recomputed the relationship between \Macc\ and the flux of the CaII866 nm line (the green 
line in Fig.~\ref{caii_comp}) considering 
the samples analyzed by Mohanty et al 2005, HH08 and in this work (ignoring the upper limits). 
We find a flatter relationship with respect to those found by the other authors (see Table~\ref{new_coeff}).  

\begin{figure}
   \centering
   \includegraphics[width=9cm]{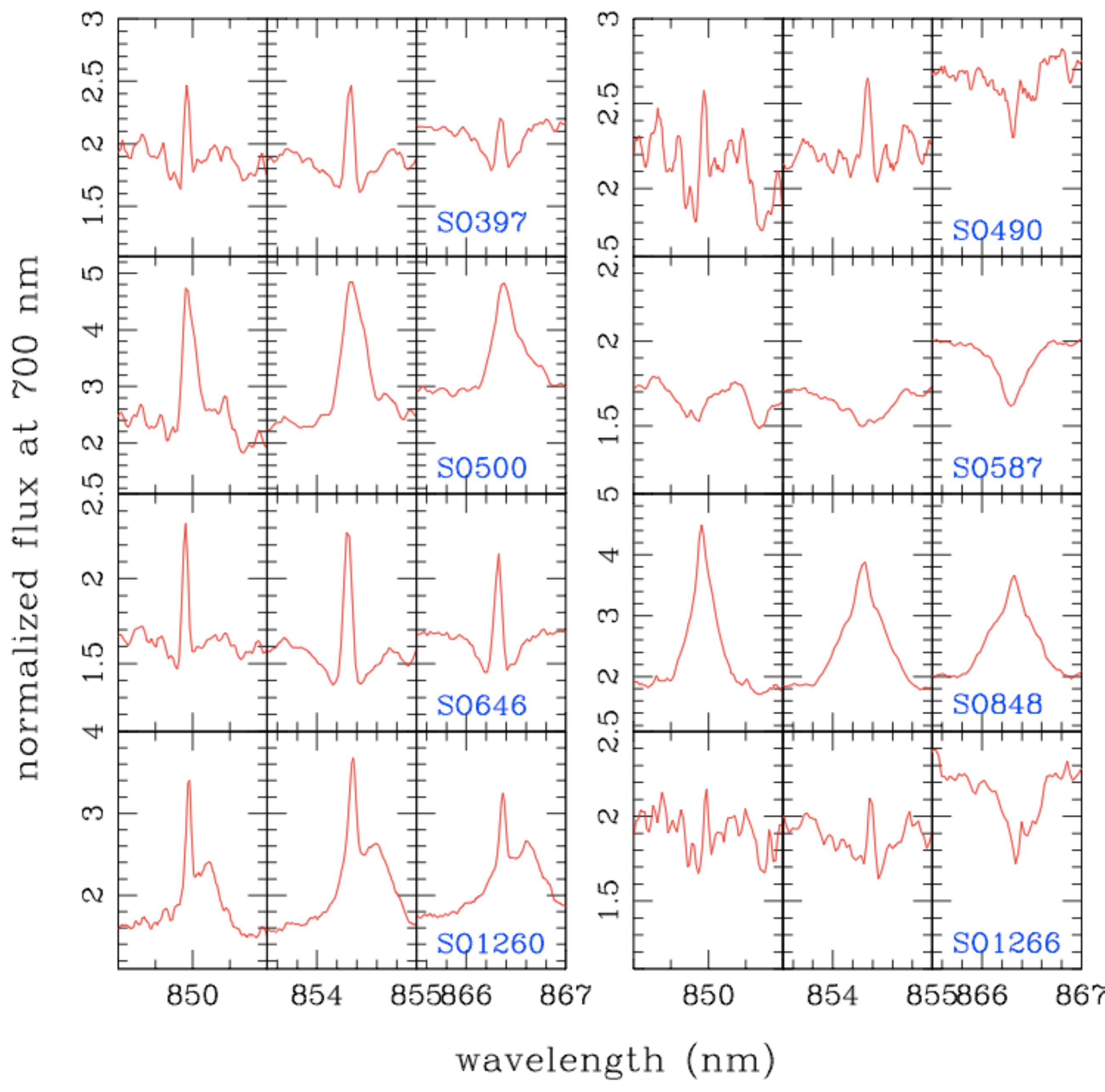}
   \caption{CaII infra-red triplet line profiles of the class II objects. For each star three panels are presented: the left panel shows the CaII$\lambda$849.8nm line, the middle panel displays the CaII$\lambda$854.2 nm line, and the CaII$\lambda$866.2 nm line is shown in the right panel.}
              \label{Ca_prof}
    \end{figure}

Even though \Macc\ and the CaII 866 flux are correlated, the spread in flux for any given \Macc\ is 
larger than two orders of magnitude, and it is not trivial  to 
decide which relation should 
be used to obtain an estimate of \Macc\ if the CaII 866 nm flux is known. 
This large spread is partly due  to the uncertainty in placing the continuum around the 
CaII Infra-Red Triplet (IRT) lines. We show in Figure~\ref{Ca_prof} the line profiles of this tracer in the 
sample analyzed here. In some cases (e.g. SO397 and SO646) the line emission 
is superposed on the absorption profile.
This could  cause an error in our estimate of the line fluxes. 
Moreover, as already noted by Mohanty et al. (2005), the CaII IRT lines are seen in 
emission only in accretors, but not all the accretors show emission in these lines. \\
New relationships, including our sample, have  also been computed 
for the CaII$\lambda$854.2 nm line and 
NaI$\lambda$589.3 nm line, and are reported in Table~\ref{new_coeff}.

\begin{figure}
   \centering
   \includegraphics[width=9cm]{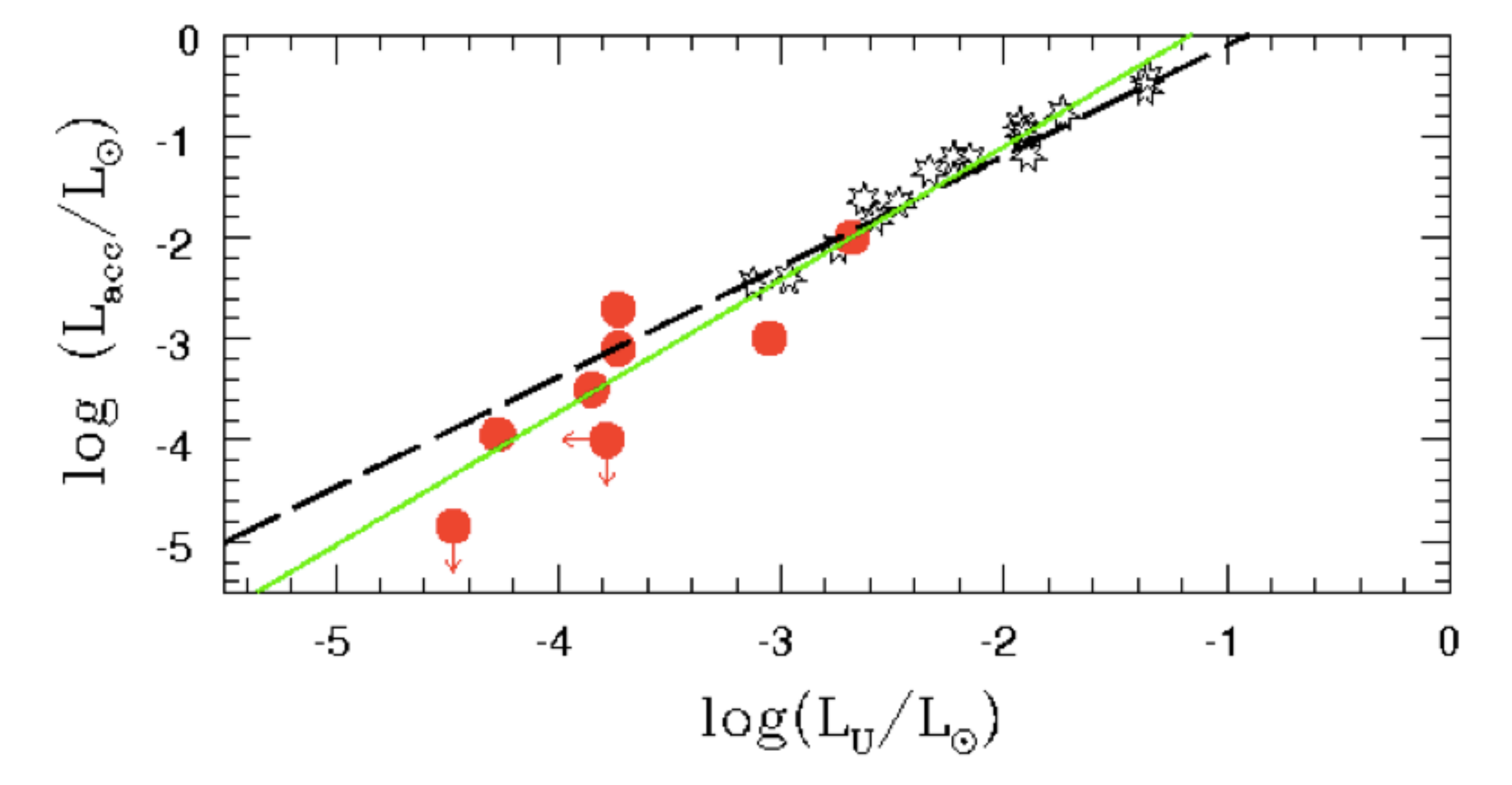}
   \caption{Relation between the U-band brightness and total excess luminosity. The  black asterisks refer to the sample analyzed by Gullbring et al.~(1998), where they found the relation drawn as dashed-black line. The red dots are the sample analyzed in this paper, and the green line is the relation found considering both the sample together. }
              \label{uband_comp}%
    \end{figure}

Fig.~\ref{uband_comp} shows the comparison between \Laccc\ and the U-band photometric 
excess (L$_U$). The U-band photometry for the X-shooter sample has been obtained by 
Rigliaco et al.~(2011a) using FORS1@VLT, and thus is not simultaneous with the other accretion diagnostics analyzed since here.
L$_U$ is highly correlated with \Laccc, as also shown by Gullbring et al.~(1998). 
With our sample we expand the relation to lower  luminosities. \\
The new relationship is slightly different with respect to the relation found 
 by Gullbring et al.~(1998), and this is clearly due to the  extension of the
sample toward lower mass accretion rate regimes. 
These results confirm that, if the extinction can be estimated, or if it is negligible, 
as in the $\sigma$ Orionis star forming region, good accretion luminosities can be 
derived from the U-band photometry, also in low accretion luminosity regimes.

\subsection{Notes on some targets}
\label{notes}

We have seen in Sect.~\ref{sect_comparison} that the accretion luminosity from the 
Balmer and Paschen excess continuum (\Laccc) is within the 1$\sigma$ uncertainty of  the 
average accretion luminosity ($\langle$\Laccl $\rangle$) 
computed considering all the secondary accretion indicators observed simultaneously, 
for five out  of eight of the targets in our sample.
Here we investigate the three sources where \Laccc\ and \Laccl\ are discrepant.\\

{\bf SO848:} \Laccc\ and $\langle$\Laccl $\rangle$  differ  by about a factor three. Looking at Fig.~\ref{Lacc_tot_average}, we can 
easily see that for SO848 the estimates of \Laccl\ from the CaII lines are the ones that 
are discrepant from all other estimates.
In the previous sections we have seen  that the CaII lines  are the  most 
uncertain  of the secondary accretion indicators, and that the estimate of the 
accretion luminosity strongly depends on the rate of accretion.  
If we neglect the accretion luminosity obtained from  the CaII lines and 
recompute the average \Laccl\, we find that the discrepancy with \Laccc\ becomes less than a factor two. 
 These differences are 
within our uncertainty in the estimate of \Laccc\ ($\langle$log\Laccl $\rangle$ neglecting the 
CaII points is $\sim$-3.21). \\

\begin{figure}
   \centering
   \includegraphics[width=9cm]{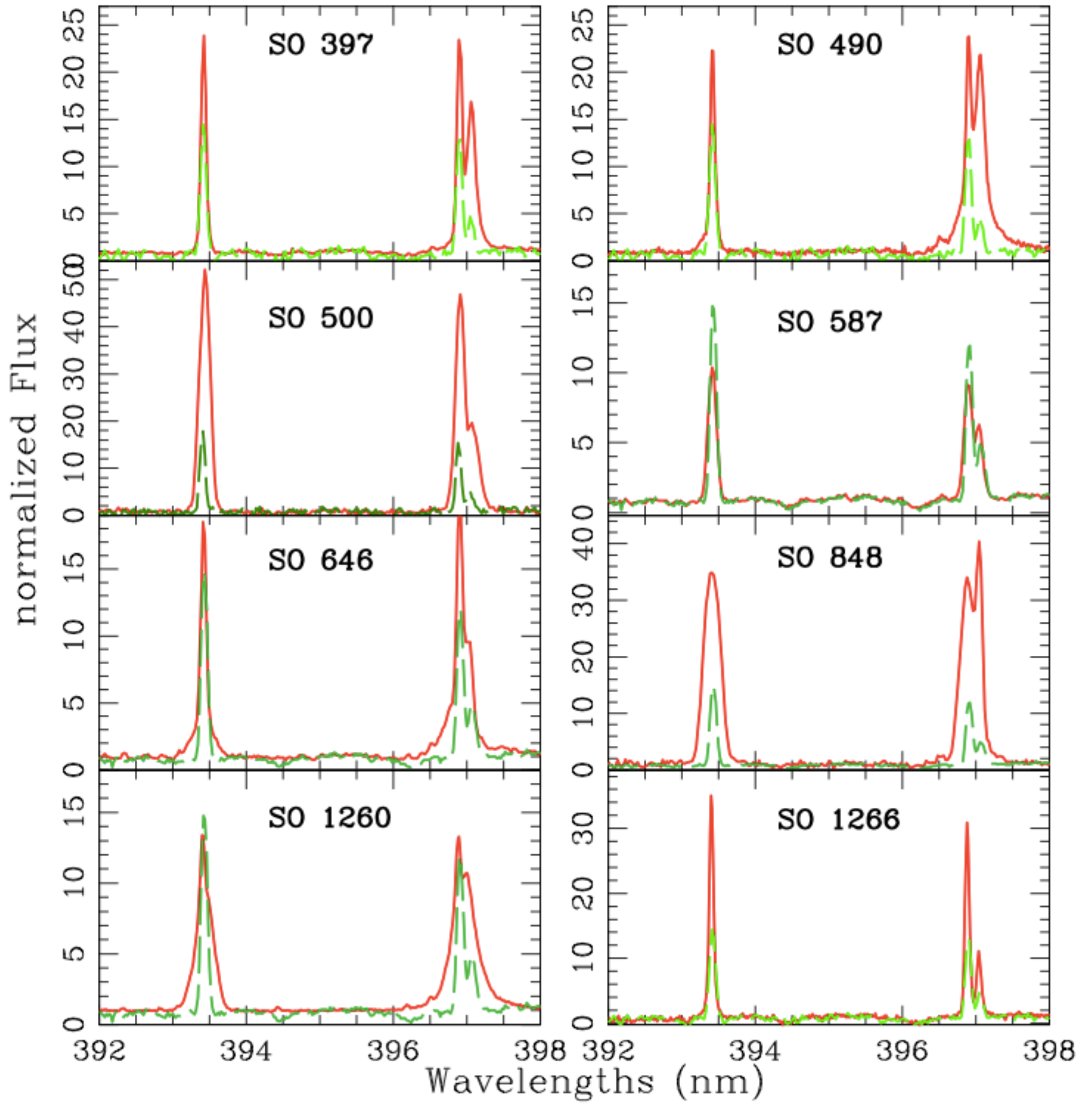}
   \caption{Each panel shows the comparison between the target stars and the templates used to obtain the continuum accretion luminosity (as listed in Table~\ref{frac_lum}) in the region of the CaII H ($\lambda$393.37 nm) \& CaII K ($\lambda$396.85 nm) lines. 
   The CaII K line is blended with the H$\epsilon$ at $\lambda$ 397.0 nm. These lines 
   are tracers of the chromospheric activity of the young stars.}
              \label{chrom_act}%
    \end{figure}

{\bf SO1266:} This object shows only a few features usable for  
deriving the accretion luminosities, and $\langle$\Laccl $\rangle$ 
is the  lowest of all values in our sample, while \Laccc\ is only an upper limit. 
Young stellar objects are generally 
characterized by strong and intense  chromospheric activity 
(see Fig.~\ref{chrom_act}). 
This activity produces also Hydrogen recombination lines, as well as the CaII H\&K 
lines, among many others. 
We find that the Balmer series line of the class III objects 
is $\sim$80\% of the line luminosity of SO1266. 
 This means that for this object, using the secondary accretion indicators, 
we are mainly tracing the  chromospheric activity of the star, with only 
$20$\,\%  of the line luminosities due to the accretion activity. 
The lack of a detectable accretion and the shape of the SED (see Fig.~\ref{SED_tot}) suggest  
that the disk surrounding SO1266 is a transitional disk, in which the inner part is completely or nearly completely 
cleared of small dust but the outer disk is optically thick (Calvet et al. 2002, Cieza 2008). 
Note that Hernandez et al. (2007) do not include this object among their transitional disk candidates. \\

{\bf SO587:} In this object  there is no evidence of excess continuum emission, and   
the estimate of \Laccc\ is an upper limit.
This object has been already analyzed by Rigliaco et al.~(2009).  
The star is most likely being photo-evaporated by the high-energy photons either from 
the central star or from the nearby hot star $\sigma$Ori. 

One  consideration to take into account in the analysis of the emission lines of this 
object, is that the photoevaporative processes could produce some emission in the 
permitted lines, which, added to the chromospheric activity of the star, could be 
wrongly interpreted as due to accretion. 
However, this hypothesis still needs to be studied in depth e.g., by analyzing higher-resolution spectra in the 
regions where lines produced by photoevaporation are expected.

\section{Summary}

We have analyzed the broad-band, medium-resolution, high-sensitivity X-Shooter 
spectra of a sample of 
young very low-mass class II and class III stars in the 
$\sigma$\,Ori star forming region.
The sample has been defined to cover the mass range 
$\sim$0.08-$\sim$0.3 \Msun.  In this paper, we focus on the accretion 
properties of the class\,II stars and the class\,III stars are used only
as spectral templates. We have carried out a comparison between 
several accretion diagnostics observed simultaneously, and spanning from the UV excess 
continuum to the IR Hydrogen recombination lines,  
computing the accretion luminosities and the mass accretion 
rates. We obtain the following results: 

1. We detect clear evidence of excess emission in the Balmer continuum in six 
 out of eight class II objects. Two of these also show hydrogen Paschen continuum emission. 
We estimate  for all eight class\,II sources  
the accretion luminosity from the excess continuum emission by 
fitting the observed continuum as the sum of the emission of a class III template and of a slab of hydrogen of 
varying density, temperature and length, under the 
assumption of LTE and without taking into account the contribution of the emission 
lines. We find that log\Laccc\ ranges between $\sim$-2.0 to  $\sim$-4.0 (\Lsun) in the objects 
where the excess continuum is measured, which converts to a mass accretion rate 
\Macc\ in the range $\sim 10^{-9}$ to $\sim 5\times 10^{-11}\,{\rm M_\odot/yr}$. 

2. We estimate the accretion luminosity from 10 secondary accretion indicators
using the  empirical relationships  with line flux and/or
luminosity given in  the literature. 
The accretion luminosities computed from all the observable tracers 
show a dispersion around the average \Laccl\ value of more than one order of magnitude 
which  can not be explained by temporal variability  given the simultaneous
observation of all accretion tracers.
This large scatter may reflect a real spread in the values of the 
secondary indicators for fixed accretion luminosity, 
owing to, e.g., different stellar and/or disk properties, or differences in wind/jet 
contributions. We investigate the possibility that such a spread may depend on the 
stellar masses or different regimes of mass accretion, and we find that these 
dependencies can be ruled out, agreeing with the results found by Curran et al. (2011) 
for higher-mass objects. 
 Despite this large spread,  for a given star all 
accretion luminosity indicators give values of 
\Laccl\ which are consistent within the  
error bars, 
and the average value of \Laccl\ has a considerably reduced uncertainty, which is about 
a factor two. 
The average accretion luminosity from the secondary accretion indicators ($\langle$log\Laccl $\rangle$) ranges between 
$\sim$-1.9 to  $\sim$-4.0 (\Lsun), which converts in \Macc\ in the range $\sim 2\times 10^{-9}$ 
to $\sim 2\times 10^{-11} $ (\Msun/yr). Moreover, $\langle$log\Laccl $\rangle$ is consistent with 
log\Laccc\ obtained from the excess continuum emission.

3. With respect to the reliability of the secondary accretion indicators, we find that 
in general the Hydrogen recombination lines, spanning from  the UV to the NIR, 
give good and consistent measurements of \Lacc\, 
often in better agreement than the uncertainties introduced by the adopted correlations. 
The average \Lacc\ derived from several Hydrogen lines, measured simultaneously, have a much reduced error.
This suggests that some of the spread in the literature correlations may be  due to the  use of non-simultaneous observations of lines and continuum, and that their quality will be significantly improved when a larger sample of X-shooter spectra will be available.

4.  We recompute the relationships  log\Lacc\ -- log$L_{line}$ for the secondary 
accretion diagnostics used  in this paper, combining  our new data with those 
available in  the literature. In most cases the relationships change  only 
within the errors but in some cases (namely log\Lacc\ -- log$L_{H\alpha}$ and log\Lacc\ -- 
log$L_{CaII}$) the correlations change significantly when we include the 
low-accretion regime sampled in this paper.
In particular, the correlation log\Lacc\ -- log$L_{H\alpha}$ shows a flattening for 
log\Lacc\ $< \sim$-3. 
This could be due to the increasingly important contribution  coming from  
chromospheric activity for weak accretors. 
The correlation log\Lacc\ -- log$L_{CaII}$ shows a very big spread in log$L_{CaII}$ 
for any given log\Lacc. 
As already noted by other authors (Muzerolle et al. 1998, Mohanty et al. 2005, HH08) 
this indicator could provide a very uncertain estimate
of \Lacc\ and/or \Macc. 
In general, a larger spread, as compared to the correlation with Hydrogen lines,
can be seen in the NaI line. 

5. The comparison between log\Laccc\ and $\langle$log\Laccl $\rangle$ obtained using the 
excess continuum emission and the emission line luminosities 
gives very encouraging results. 
The excess  Balmer continuum accretion luminosities are in very good agreement 
with the average $\langle$log\Laccl $\rangle$ in five out of eight stars of our 
sample. The remaining three objects do not show such agreement. 
We discussed these three cases separately, finding that in each case the disagreement 
can be attributed to peculiar characteristics of the targets (strong wind due to 
photoevaporation processes acting on  SO587, wrong interpretation of the 
 chromospheric activity for SO1266, and the large uncertainty of the CaII lines as 
accretion tracers for SO848). 
We conclude that, the average accretion luminosity computed as the average of several 
secondary accretion indicators is as reliable as the accretion luminosity obtained 
from a the excess continuum emission, except for peculiar cases.

\begin{acknowledgements}
     E.R. thanks Ilaria Pascucci for valuable discussions. The authors acknowledge C. Manara for helping with the class III targets. The authors are thankful to the ESO staff, 
in particular C. Martayan for support in the observations, and P. Goldoni and A. Modigliani for 
their help with the X-shooter pipeline. 
\end{acknowledgements}

\end{document}